\documentclass[fleqn,10pt]{wlscirep}
\usepackage[utf8]{inputenc}
\usepackage[T1]{fontenc}

\usepackage{graphicx}
\usepackage{dcolumn}
\usepackage{bm}
\usepackage[colorlinks=true, allcolors=blue]{hyperref}
\usepackage{amsmath}
\usepackage{physics}
\usepackage{amsfonts}
\usepackage{amssymb} 
\usepackage{braket}
\usepackage{comment}
\usepackage{subfig} 

\usepackage{algorithm}
\usepackage{algpseudocode}

\usepackage[numbers,square,comma,sort,compress]{natbib}
\usepackage{chapterbib}

\title{Measurement-Based Quantum Compiling via Gauge Invariance}

\author[1, 3]{Sebastiano Corli}
\author[2, 3, *]{Enrico Prati}
\affil[1]{Dipartimento di Fisica, Politecnico di Milano, Piazza Leonardo da Vinci 32, 20133 Milano, Italy}
\affil[2]{Università degli Studi di Milano, Dipartimento di Fisica "Aldo Pontremoli", via Celoria 16, 20133 Milano, Italy}
\affil[3]{Istituto di Fotonica e Nanotecnologie, Consiglio Nazionale delle Ricerche, Piazza Leonardo da Vinci 32, 20133 Milano, Italy}

\affil[*]{enrico.prati@unimi.it}

\begin{abstract}
The measurement-based architecture is a paradigm of quantum computing, relying on the entanglement of a cluster of qubits and the measurements of a subset of it, conditioning the state of the unmeasured output qubits.
While methods to map the gate model circuits into the measurement-based are already available via intermediate steps, we introduce a new paradigm for quantum compiling directly converting any quantum circuit to a class of graph states, independently from its size. Such method relies on the stabilizer formalism to describe the register of the input qubits.
An equivalence class between graph states able to implement the same circuit is defined, giving rise to a gauge freedom when compiling in the MBQC frame.
The graph state can be rebuilt from the circuit and the input by employing a set of graphical rules similar to the Feynman's ones.
A system of equations describes the overall process.
Compared to Measurement Calculus, the ancillary qubits are reduced by 50\% on QFT and 75\% on QAOA.
\end{abstract}

\begin{document}

\flushbottom
\maketitle
%
%
\thispagestyle{empty}



\begingroup
\let\clearpage\relax

\citestyle{nature}
\bibliographystyle{unsrtnat}

\section*{Introduction}


The measurement-based quantum computing (MBQC) also referred to as one-way quantum computing \cite{raussendorf2001one} has become a viable alternative computational paradigm to gate model quantum computing mainly thanks to the development of photonic quantum computers \cite{wang2020integrated}. Instead of applying a universal set of unitary operators to a register of initialized qubits, the information processing is carried out after entangling the qubits in a source state~\cite{dunjko2012blind,kissinger2019universal}. Afterwards, measuring a subset of the source state, the quantum state of the unmeasured ones (the output qubits) is purposely altered~\cite{joo2019logical,scott2022timing,proietti2022native}.  The irreversibility of the measurement justifies the alternative name of one-way quantum computing.
In the MBQC frame, a major hindrance consists of entangling all the qubits into a cluster.
The main hurdles, when entangling a cluster of qubits, raise from technological limitations, e.g. photon loss~\cite{scott2022timing, benjamin2006brokered}. However, protocols such as entanglement purification, or entanglement witnesses, allow to tackle such withdrawals, achieving a high fidelity on the source states~\cite{zwerger2014hybrid,kruszynska2006entanglement,dur2003multiparticle,aschauer2005multiparticle,glancy2006entanglement,bera2022class}.
Next to the generation of the source state, two-qubits gates are realized by single-qubit measurements during the computation~\cite{menicucci2014fault,zwerger2014hybrid}.
Therefore, once the entangled state has been allocated, it is possible to substitute coherent and long-ranged operations on qubits, envisaging MBQC as a viable architecture in terms of qubit-scalability and fault-tolerance~\cite{benjamin2006brokered,larsen2019deterministic,raussendorf2007topological}.
In MBQC, the entangling gates are applied just before the algorithm executes, thus any error deriving from such process will alter the cluster of entangled qubits, but not necessarily the algorithm itself, which makes a subtle difference with the error correction and mitigation techniques developed for the gate model~\cite{benjamin2006brokered}.
Indeed, the latter assumption drops the need to maintain coherent control over the quantum state of qubits, replacing it with the less experimentally challenging goal of initializing a graph state before starting the computation~\cite{mantri2017universality}.
%
%
%
%
%
In their proposal, Danos et al.~\citep{danos2007measurement,pius2010automatic} provided a method, the MCalculus, to translate a sequence of logic gates into a set of entangling gates, quantum measurements and Pauli corrections to perform in order to translate the gate sequence into the MBQC formalism. Such sequence of operations is named the MQBC pattern~\cite{pius2010automatic}.
Other methods to convert unitary operations into graph states have been proposed, even without implementing a universal computation, for example to convert diagonal operators into specific graphs~\cite{browne2016one}. Recently, Wong et al.~\cite{wong2024gauge} formulated an underlying gauge theory over the MBQC, contextualized in the frame of condensed matter and high-energy physics, not related to quantum compiling.
%
Previously, we have already addressed graph state encoding~\cite{corli2022efficient} and gate-model related quantum compiling \cite{maronese2022quantum} by a deep learning approach for both single gate \cite{semola2022deep} and unitary approximation \cite{moro2021quantum}, respectively. 
%
%
Instead, here we develop an exact method for MBQC compiling, able to map any quantum circuit into a graph state without any transposition from the gate model.
The description of the qubit system takes advantage of the notation previously introduced in Ref.~\cite{corli2022efficient} and it is described by group theory instead of state vectors or the density matrices. Indeed, the overall state is represented by a set of stabilizers which act on the initialized register $\ket{0}^{\otimes n}$.
Such formalism relies on applying a decomposition of the unitary operator in terms of Pauli matrices, whose goal can be tackled via the KAK decomposition~\cite{khaneja2001cartan,saravanan2021decomposition,rodriguez2012kak,lao2022software,bullock2004canonical}, or rather by any decomposition of a generic unitary in terms of Pauli matrices. The decomposition of the circuit into a set of universal gates could be employed as well, by opportunely handling the rules of Clifford group introduced in the next Sections.
%
%
Thanks to such decompositions, we prove that any unitary operator can be converted directly into a graph state, bypassing any circuit conversion from gate-based to measurement-based formalism.
%
Indeed, we introduce a straightforward construction of the graph relying on a technique which we call \textit{analysis of the residual}. Such technique relies on a set of graphical rules, resembling the Feynman's ones, which allow to convert a graph state into a linear combination of operators and vice versa.
Furthermore, we show that multiple graph states, belonging to an equivalence class, are able to apply the same quantum circuit to a specific input state. Such equivalence gives rise to a gauge freedom, as different graph states can be chosen without affecting the outcome of the computation.
We prove that, in order to construct the graph, for mapping any generic unitary $\hat U$ representing the quantum circuit into a graph $G[V,E]$, at least one explicit graph representing the solution exists. We refer to this mapping as the fully symmetric gauge.
Such gauge freedom, as well as the overall method, is expressed by a system of algebraic equations.
Exploiting the fully symmetric gauge, we cut the number of ancillary qubits, with respect to the M-Calculus, by a factor of $1/4$ for QAOA and $1/2$ for QFT.
Moreover, such formulation in terms of equations allows to invert the already acknowledged methods for mapping unitary operators into a graph state (as the MCalculus).


%
\begin{figure}
\subfloat[\label{subfig:MBQC}]{{\includegraphics[width=0.95\textwidth]{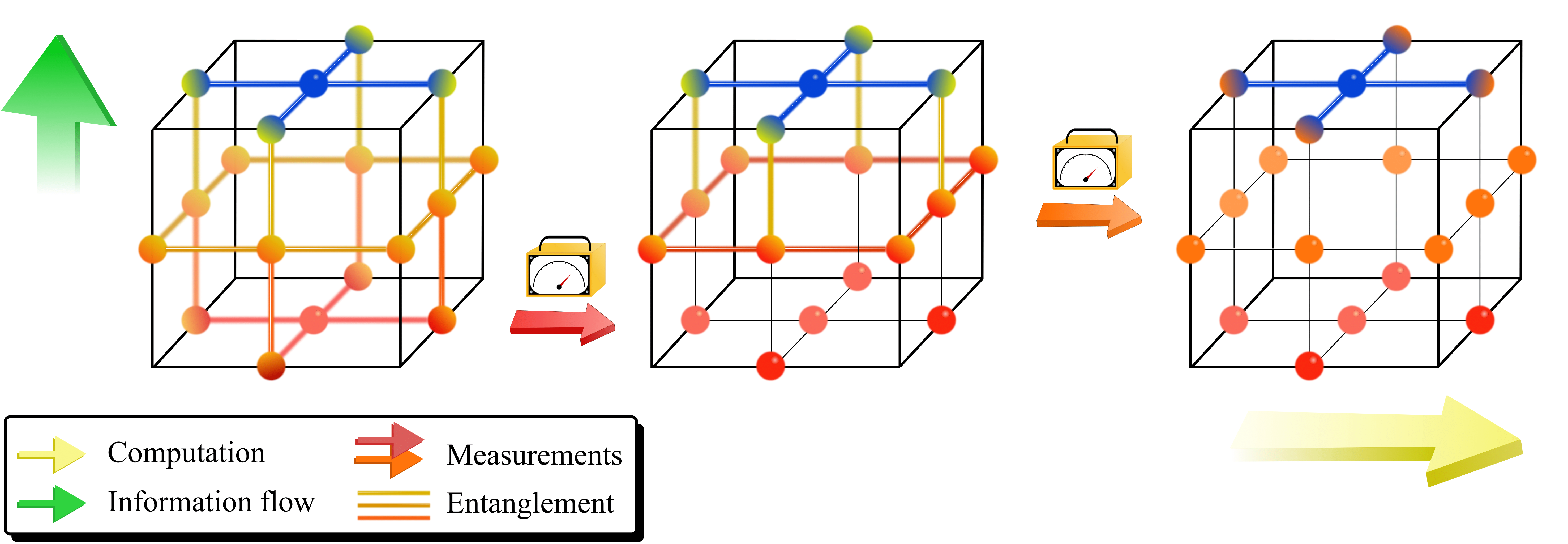} }}
\\
\subfloat[\label{subfig:models}]{{\includegraphics[width=0.95\textwidth]{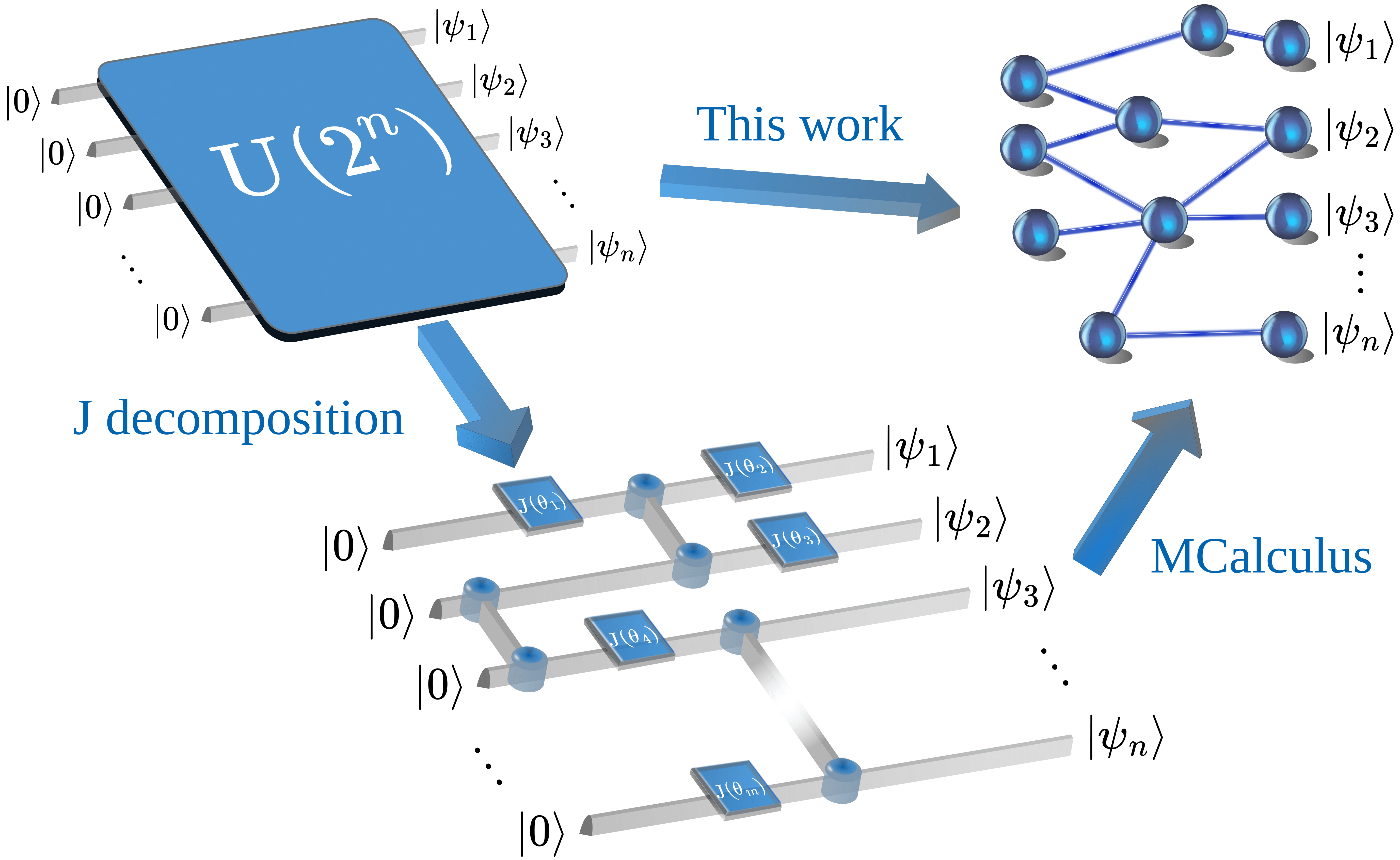} }}
\caption{(a) In the MBQC paradigm, the graph state (here represented by two nodes linked by the violet line, representing their entanglement) is split into the source state, i.e. the qubits which have to be measured to let the information flow (in a blue and red superposition) and the output states (red and blue superposition). Due to the entanglement, the output qubits are affected by the measurements over the source qubits. (b) In quantum compiling, a generic unitary operator needs to be decomposed into elementary operations. 
In the measurement-based architecture, the unitary operation is prompted by entangling the source states with the output qubits, thereafter measuring the previous ones. Currently, the process consists of two steps, namely J decomposition of the unitary operator and next the construction of the graph state by using MCalculus. The aim of this work is to build a compiling algorithm to cast directly the unitary matrix from $\hat U(2^n)$ into the graph state.
}
\label{fig:MBQCframe}
\end{figure}

\section*{Results}

The MBQC paradigm consists of entangling a cluster of qubits and progressively measuring a subset of source qubits, whose entanglement with the unmeasured qubits conditions the output state of the latter ones (Figure \ref{subfig:MBQC}).
The rational behind the direct compiling method, which prepares the starting graph state from any given arbitrary unitary operator -- representing a quantum circuit -- discussed in this Section is represented in Figure \ref{subfig:models}.
The demonstration of the validity of this compiling method is demonstrated by construction through three major steps. The first step consists of the description of the input state, by merging the circuit and the state of the register by a single operator, later called unitary supplement. The second step consists of preparing the equivalence class which leaves the output state unchanged after having consistently altered both the input circuit operator and the input state (expressed in turn as an operator). Finally, as a third step, such gauge invariance provides the MBQC embedding by graphical rules, as happens in Feynman rules in perturbative quantum field theory, defining an equivalence class of graphs, from which the MBQC algorithm starts to be computed. The overall picture of the MBQC compiling is expressed by a system of core equations.

\subsection*{Preparation of the input state}

Le't start by describing the first step, consisting of merging the given circuit with the register of input qubits.
First, the input circuit $\hat U$ is expressed by its decomposition into rotations induced by elements of the Pauli algebra.
Next, the input state is
expressed by operators $\hat G_i$ called generators, $i$ referring to each qubit, instead of the usual vector state.
Eventually, the composition of the circuit $\hat U$ with the generators $\prod_i \hat G_i$ gives rise to the unitary supplement $\eta$, a tensor representation of the original circuit and input register by a single operator $\hat U \prod_i \hat G_i$.
\newline
\newline
\textbf{Pauli decomposition.} The input circuit $\hat U$ of $n$ qubits is first represented by rotations induced by elements from the Pauli group $\hat \sigma \in \mathcal{P}$. Such representation can be borrowed e.g. from the KAK decomposition:
\begin{equation}
\label{eq:Udecomp}
   \hat U = \hat U_l \circ \hat U_{l-1} \circ ... \circ \hat U_1 = [\cos(\theta_{l}) + i \sin(\theta_l) \sigma_1] \circ ... \circ [\cos(\theta_1) + i \sin(\theta_1) \sigma_1]
\end{equation}
Each of the $\hat U_i$ refers to a specific operation described in a quantum circuit, which may involve a number of qubits ranging from $1$ to $n$.
\newline
\newline
\textbf{Expression of the input state as generators.}
A suitable formalism to describe the input state is introduced. 
Given a register $\ket{\Psi}$ of $n$ qubits, $\ket{\Psi} \in [\mathbb C^2]^{\otimes n}$, the \textit{generator} $\hat G_i$ of the $i$-th qubit is defined as a linear combination of the identity $\hat I_{N\times N}$ ($N=2^n$) plus a Pauli operator $\hat K_i = \hat X_i \bigotimes_{j=1}^n \hat \sigma_j$ which we call \textit{stabilizer}, where $\hat \sigma_j \in \{ \hat I_{2 \times 2}, \hat Z\}$, the latter being a subgroup of the Pauli group $\mathcal{P}_1$.
\begin{equation}
    \hat G_i(a,b) = a \hat I_{N \times N} + b \hat K_i
\end{equation}
From now on, for sake of ease, $\hat I_{N \times N}$ will be omitted from the notation. In a dedicated lemma from Supplementary Note 1, we prove that the generators $\hat G$ are not unitary, even though returning $\bra{0} \hat G^\dagger \hat G \ket{0} = 1$. Any quantum state encoding a single qubit can be represented by
\begin{equation}
\label{eq:qubitInstalled}
    \ket{\psi} = a \ket{0} + b \ket{1} = [a + b \hat X] \ket{0} = \hat G(a,b) \ket{0}
\end{equation}
A register of $n$ non-entangled qubits can be generalized as
\begin{equation}
\label{eq:MultipartState}
    \ket{\Psi} = \prod_{i=1}^n [a_i + b_i \hat X_i] \ket{0}^{\otimes n} = \prod_{i=1}^n \hat G_i(a_i, b_i) \ket{0}^{\otimes n}
\end{equation}
The coefficients in front of the $\hat K_i$ operator (the $b_i$ ones) are from now referred to as the \textit{interactive coefficients} of the generator $\hat G_i$. The coefficients in front of the $\hat I_{N \times N}$ identity (the $a_i$ ones) as the  \textit{idle coefficients} of the $\hat G_i$ generator. The entanglement between two generators $\hat G_i$, $\hat G_j$ can be induced by the $\hat{CZ}_{ij}$ phase operator, morphing their stabilizer $\hat X_i \overset{\hat{CZ}_{ij}}{\to} \hat X_i \hat Z_j = \hat K_i$. A qubit $i$ entangled with a set $\mathcal{N}(i)$ of qubits displays a stabilizer like
\begin{equation}
\label{eq:XtoK}
    \hat K_i = \hat X_i \bigotimes_{j \in \mathcal{N}(i)} \hat Z_j
\end{equation}
$\mathcal{N}(i)$ representing the sequence of all the $\hat{CZ}_{ij}$ Clifford operations.
In Supplementary Note 2, we prove how the $\hat K_i$, $\hat K_j$ stabilizers still commute each other with, even after applying the $\hat{CZ}_{ij}$ phase gates.
In Supplementary Note 3, we summarize the rules for the Clifford algebra, which $\hat{CZ}_{ij}$ belongs to.
Therefore, when dealing with MBQC, the generators allow to implement the overall system in a much more compact representation then by installing a state vector.
\newline
\newline
\textbf{Construction of the unitary supplement tensor.}
We define the \textit{unitary supplement} $\hat \eta$ as the composition between the product of the generators $\prod_i \hat G_i$ and the unitary implementing the quantum circuit:
\begin{equation}
\label{eq:QSystem}
    \hat \eta = \hat U \circ \prod_{i=1}^n \hat G_i
\end{equation}
The unitary supplement, as a  $2^n \times 2^n$ matrix, can be expressed in a tensor formalism by the canonical basis $\hat e_i$ of the $2 \times 2$ matrices:
\begin{equation}
\hat \eta = \eta^{i_1 ... i_n} \hat e_{i_1} \otimes \hat e_{i_2} \otimes ... \otimes \hat e_{i_n}, \qquad \hat e_i
\in \{\begin{pmatrix}
    1 & 0 \\ 0 & 0
\end{pmatrix}, \begin{pmatrix}
    0 & 1 \\ 0 & 0
\end{pmatrix},
\begin{pmatrix}
    0 & 0 \\ 1 & 0
\end{pmatrix},
\begin{pmatrix}
    0 & 0 \\ 0 & 1
\end{pmatrix}\}    
\end{equation}
where we followed the Einstein notation on the indices.

\subsection*{Definition of the equivalence class}

\begin{figure}
\centering
\includegraphics[width=0.8\textwidth]{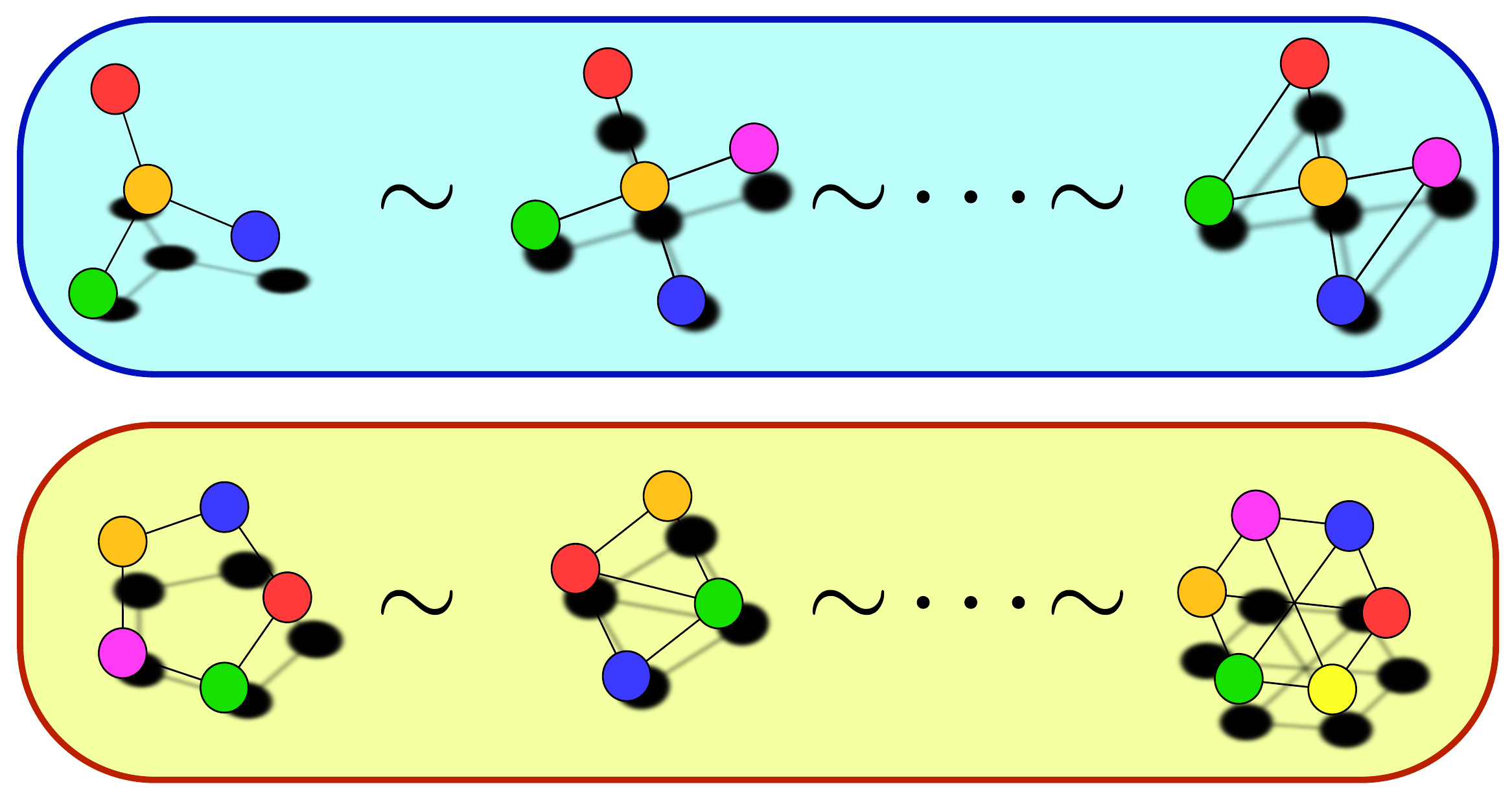}
\caption{Graphical representation of two equivalence classes. All of the elements inside a class share a common $\vec \eta$ vector supplement. The maps between graphs belonging to the same class are implemented by the $S$ gauge tensor.}
\label{fig:SecondTakeCX}
\centering
\end{figure}

Once the input register and the quantum circuit are unified into the unitary supplement tensor, we now turn to the definition of equivalence class of the unitary supplement operators. 
In order to natively adopt operatorial expressions ready for the next step of the compiling process, we aim to describe the quantum circuit and the register as a linear combination of elements from the Pauli algebra. Therefore, in order to the define the equivalence class, we start by expressing the unitary supplement tensor in the Pauli basis.
Next, the equivalence class is established between those combinations of quantum circuits $\hat U$ and the register $\prod_i \hat G_i$ which act identically over $\ket{0}^{\otimes n}$, i.e. which are characterized by a unitary supplement with identical first column. Eventually,
we introduce a map between elements of the same equivalence class, which 
constitutes the gauge transformations exploited in the final step of the compiling method. 
\newline
\newline
\textbf{The unitary supplement in the Pauli basis.} Due to Equations \eqref{eq:Udecomp} and \eqref{eq:MultipartState}, the unitary supplement $\hat \eta$ can also be described as a linear combination with generators from the Pauli group $\sigma \in \{\hat I, \hat X, \hat Z, \hat Z \hat X \}$. Such linear combination in the Pauli basis is referred to as the \textit{transfer tensor} $\hat T$:
\begin{equation}
    \hat T = T^{i_1... i_n} \hat \sigma_{i_1} \otimes \hat \sigma_{i_2} \otimes ... \otimes \hat \sigma_{i_n}
\end{equation}
The transfer tensor and the unitary supplement describe both the same tensor object but for a change of basis, implemented by an isomorphism $M$:
\begin{equation}
\label{eq:Isom}
    \eta_{i_1 ... i_n} = M_{i_1}^{\; j_1} ... M_{i_n}^{\; j_n} T_{j_1... j_n}
\end{equation}
\newline
\newline
\textbf{Criterion to establish the equivalence class.} By construction, the action of the unitary supplement is expressed over the $\ket{0}^{\otimes n}$ state:
\begin{equation}
\label{eq:SuppVec}
    \hat \eta \ket{0}^{\otimes n} = \vec \eta
\end{equation}
where $\vec \eta$ is the first column of the $2^n \times 2^n$ matrix implemented by $\hat \eta$. From now on, $\vec\eta$ is referred to as \textit{vector supplement}.
\newline
\newline
\textbf{Equivalence class between circuits.} As the output of the computation is determined by the sole $\vec \eta$ vector supplement, as stated by Equation \eqref{eq:SuppVec}, a class of equivalence between different unitary supplements $\eta$, $\eta'$ is established, once the criterion of equivalence is set by the condition $\vec \eta = \vec \eta'$:
\begin{equation}
    \hat \eta \sim \hat \eta' \Leftrightarrow \vec\eta=\vec\eta'
\end{equation}
Transforming the unitary supplement, and therefore the circuit itself, without spoiling the output of the computation gives rise to a gauge transformation, i.e. a map between different elements of the aforementioned equivalence class. As we are interested in quantum computation, expressed by Pauli operators, we implement such gauge transformation on the transfer tensor:
\begin{equation}
\label{eq:Gaugetransf}
    T'_{i_1... i_n} = S_{i_1 ... i_n}^{\qquad j_1 ... j_n} T_{j_1 ... j_n}
\end{equation}
or introducing a shorter notation:
\begin{equation}
    T'_{\textbf{i}} = S_{\textbf{i}\textbf{j}} T_{\textbf{j}}, \qquad
    M_{i_1}^{\; j_1} ... M_{i_n}^{\; j_n} = M_{\textbf{i}\textbf{j}}
\end{equation}
We refer to the tensor $S$ as the \textit{gauge tensor} (Figure 2). Collecting all the information from Equations \eqref{eq:Isom}, \eqref{eq:SuppVec} and \eqref{eq:Gaugetransf}, the following system holds:
\begin{subequations}
\label{eq:FirstSystem}
\begin{align}
        T'_{\textbf{i}} = S_{\textbf{i}\textbf{j}} T_{\textbf{j}} = S_{\textbf{i}\textbf{j}} M_{\textbf{j}\textbf{k}} \,\eta_{\textbf{k}}, \label{subeq:T}     \\
        \vec\eta = \vec\eta', \label{subeq:VecSuppl} \\
        \eta'_{\textbf{i}} = M^{-1}_{\textbf{i}\textbf{j}} T'_{\textbf{j}} = M^{-1}_{\textbf{i}\textbf{j}} S_{\textbf{j}\textbf{k}} T_{\textbf{k}} \label{subeq:unitsupp}
\end{align}
\end{subequations}

\subsection*{MBQC graphic rules and core Equations}

\begin{figure}
\centering
\includegraphics[width=0.6\textwidth]{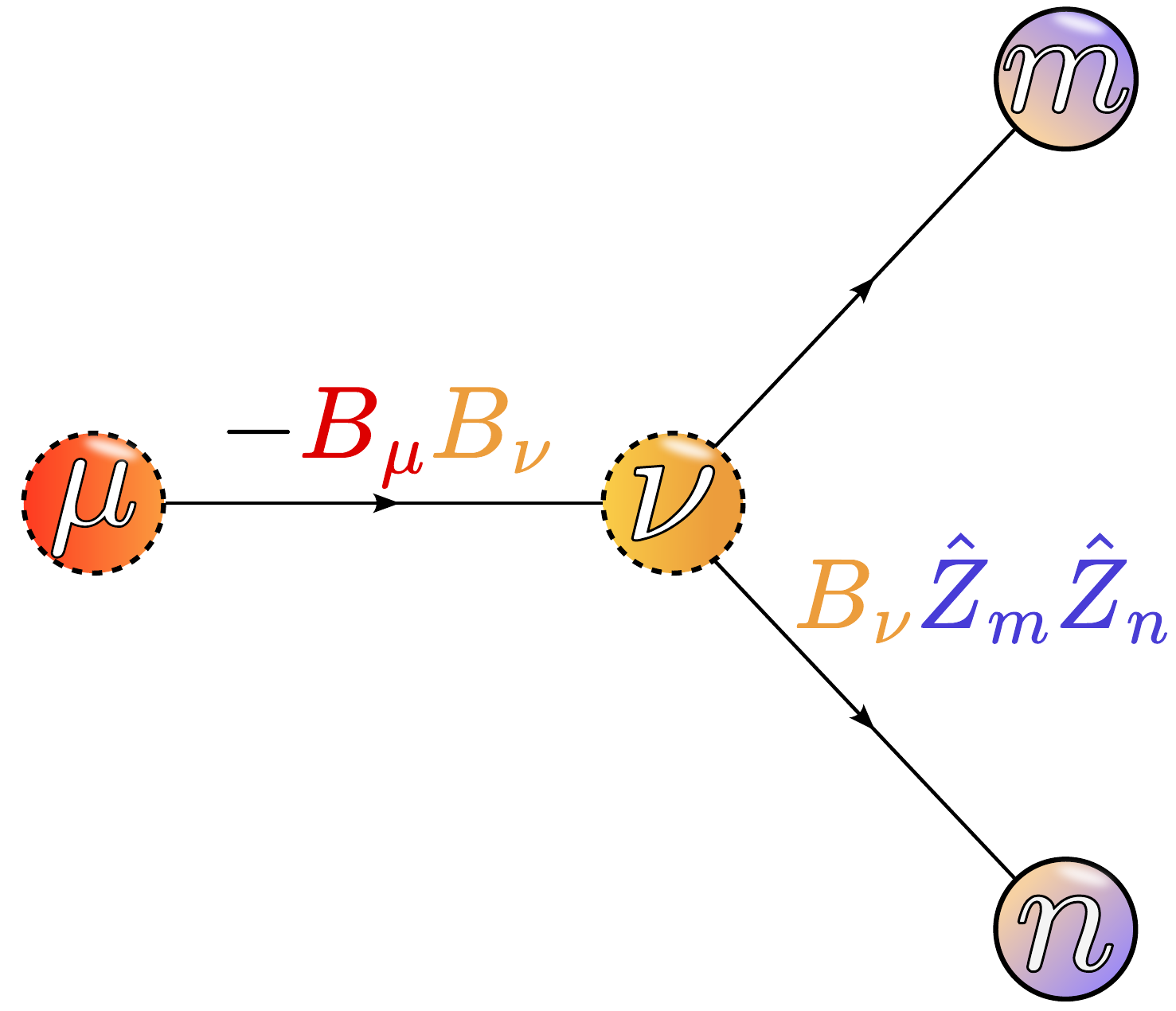}
\caption{Graphical rules which link the residual operator to the graph states. When two source qubits neighbor (greek letters, dashed borders), a minus sign occurs in front of their interactive coefficients, while a $\hat Z$ operator when a measured qubit neighbors with the output $m$, $n$ ones. The computational flow follows the arrows on the edges of the graph.}
\label{fig:Feynman}
\centering
\end{figure}

After the equivalence class is established, we explicitly refer to MBQC framework by imposing that the linear combination of Pauli operators (expressed by the transfer tensor) is the result of a measurement process over a graph state. Therefore, we build a special transfer tensor, suitable for MBQC, implementing the action of the measurement process by defining a to-be-called residual operator. Next, we apply topological rules expressed by two major theorems, similar to the Feynman's. Next, symmetry constraints are imposed to select those transfer tensors suitable for MBQC compiling and thus generating a graph state, associated to the original input circuit. This MBQC compiling method is embodied by a system of core equations.
\newline
\newline
\textbf{Measurements in the generator formalism.}
A measurement on the $i$-th qubit is implemented by a projector operator $\hat P_\gamma = \ket{\gamma_i}\bra{\gamma_i}$, where $\ket{\gamma_i}$ is the eigenstate of an observable $\hat \sigma$. After the measurement of the $i$-th qubit is performed, the measured qubit lies in the $\ket{\gamma}_i$ state. To relieve the notation, we write down the measurements as $\bra{\gamma_i}$, instead of $\ket{\gamma_i} \bra{\gamma_i}$.
A measurement over a generic state $\bra{\gamma_i}$ can be represented as
\begin{equation}
\label{eq:reprMeas}
    \bra{\gamma_i} = \alpha_i^* \bra{0_i} + \beta_i^* \bra{1_i} = \bra{0_i} [ \alpha_i^* + \beta_i^* \hat{X}_i] = \bra{0_i} \Gamma(\alpha_i, \beta_i)
\end{equation}
$\Gamma$ being the generator for such measurement, and $\alpha$, $\beta \in \mathbb{C}$, $\left|\alpha\right|^2 + \left|\beta\right|^2 = 1$.
$\Gamma$ involves no $\hat Z$ Pauli operator as the measurements act as projections over a single-qubit Hilbert space, with no entanglement involved. In literature, the measurements with respect to the Pauli observables can be performed over three planes, the $XY$, the $XZ$ and the $YZ$~\cite{browne2007generalized, joo2019logical,markham2014entanglement}, each of them being a linear combination of two Pauli matrices, whose coefficients are given as $\cos(\theta)$, $\sin(\theta)$. In Supplementary Note 4, we explain how to represent such projections in the generator representation.
\newline
\newline
\textbf{The residual operator.} The action of the measurements over an input state $\ket{\Psi}$ can be therefore described by an operator $\hat R$, which we call the \textit{residual} of the measurements, or simply the residual. Such operator is nothing but the generator of a qubit transformed after the measurement itself, i.e. after projecting the single-qubit state described by the generator over a subspace as in Equation \eqref{eq:reprMeas}.
Given a state $\ket{\Psi}$ of $n+m$ qubits, the residual $\hat R$ is the operator acting over the new $n$-qubits state $\ket{\Psi'}$, after a subset of $m$ qubits has been measured, according to:
\begin{subequations}
\begin{align}
        ^{m\otimes}\bra{\gamma_m} \bigotimes_{j=1}^m [a_j + b_j \hat K_j] \bigotimes_{l=m+1}^{n+m} [a_l + b_l \hat K_l] \ket{0}^{\otimes (n+m)}
    = \hat R \; \bigotimes_{l=m+1}^{n+m} [a_l + b_l \hat K'_l] \ket{0}^{\otimes n} \\
    \hat R = \; ^{m\otimes}\bra{\gamma_m} \bigotimes_{j=1}^m [a_j + b_j \hat K_j] \ket{0}^{\otimes m} \label{subeq:Residual2}
\end{align}
\end{subequations}
where the $\bra{\gamma_i} = \bra{0} [\alpha_i^* + \beta_i^* \hat{X}_i]$ are the measurements over the $\ket{\Psi}$ state, the latter one being allocated by the $n+m$ generators. After the measurements, the new $\ket{\Psi'}$ state is defined by $n$ new $\hat K'_l$ generators, determined by the previous measurements, while the previous $m$ generators $\hat K_j$ are transformed into the $\hat R$ residual.
In physical terms, the transformation $\hat K_l \to K'_l$ means that the entanglement between the unmeasured qubits and the measured ones has been severed. In fact, the $n$ stabilizers $\hat K_l$ of the unmeasured qubits transform as follows:
\begin{equation}
\label{eq:GenUnmeas}
    \hat K_l = \hat X_l \bigotimes_{j \in \mathcal{N}(l)} \hat Z_j \to
    \hat K'_l = \hat X_l \bigotimes_{j \in \mathcal{N}(l) \setminus \mathcal{M}} \hat Z_j
\end{equation}
where $\mathcal{N}(l)$ denotes the set of the qubits entangled with the $l$-th one before the measurements, and $\mathcal{M}$ the set of measured qubits. In order to tackle the expression of the residual in Equation \eqref{subeq:Residual2}, all the $\ket{0}^{\otimes m}$ kets must be contracted to compute the expectation value. As $\hat Z \ket{0} = \ket{0}$, the action of such contraction over the generators can be summed up as follows: suppose $p$ to be the qubit to be measured, any $\hat G_l$ generator thus transforms into a $\hat G_l'$ as
\begin{equation}
\begin{split}
    \hat G_l \ket{0_p} = [a_l + b_l \hat X_l \bigotimes_{j \in \mathcal{N}(l)} \hat Z_j] \ket{0_p} =
    \ket{0_p} [a_l + b_l \hat X_l \bigotimes_{j \in \mathcal{N}(l) \setminus \{p\}} \hat Z_j] = \ket{0_p} \hat G'_l
\end{split}
\end{equation}
proving the validness of Equation \eqref{eq:GenUnmeas}. Furthermore, the residual $\hat R$, over a register of $n$ qubits, is expressed as a linear combination of $\bigotimes_{j=1}^n \hat \kappa_j$ operators, with $\hat \kappa_j \in \{\hat I_{2\times 2}, \hat Z\}$. The residual is yielded by the following expectation value over the $i$-th qubit:
\begin{eqnarray}
\label{eq:Res1}
    \hat R = \bra{0_i} [ \alpha_i^* + \beta_i^* \hat{X}_i] [a_i + b_i \hat{X}_i \bigotimes_{j \in \mathcal{N}(i)} \hat Z_j] \ket{0_i} = \alpha_i^* a_i + \beta_i^* b_i \bigotimes_{j \in \mathcal{N}(i)} \hat Z_j = A_i + B_i \bigotimes_{j \in \mathcal{N}(i)} \hat Z_j
\end{eqnarray}
To prove such statement, refer to Supplementary Note 5.
Consequently, in the MBQC frame the action of a unitary operator can be expressed into a linear combination of $A_i + B_i\bigotimes_{j = 1}^n \hat \kappa_j$ operators. For the sake of clarity, we may now introduce two definitions.
Any coefficient $A_i \neq 0$ from the residual $\hat R$, as in Equation \eqref{eq:Res1}, is labelled as \textit{idle coefficient}. Instead, any coefficient $B_i$ from the residual $\hat R$, as in Equation \eqref{eq:Res1} and $\mathcal{N}(i) \neq \emptyset$, is called \textit{interactive coefficient}.
\newline
\newline
It is now possible to enunciate the two theorems mentioned above. The first theorem states that, after a measurement, the interactive coefficients from the generators are multiplied in the interactive coefficient of the residual, and separately the idle coefficients from the generators are multiplied in the idle coefficient of the residual, with no cross-terms. In the second theorem, we state how two measured qubits, previously entangled, output a minus sign multiplying their interactive coefficients in the residual, a property which is a key element of the compiling algorithm. Indeed, from the second Theorem it follows a Corollary which is exploited, along with the corresponding Theorem, to build the compiling algorithm for MBQC.
\newline
\newline
\textbf{Coefficients Theorem}: after a $\bra{\gamma_i}$ measurement, $\bra{\gamma_i}$ in the form of Equations \eqref{eq:reprMeas} and $i \in \{0,...,n\}$, the interaction coefficients $b_i$, $\beta_i^*$ and the idle coefficients $a_i$, $\alpha^*_i$, where $\alpha_i$, $\beta_i$ are from the generator $\hat \Gamma$ of the measurement and $a_i$, $b_i$ from the generator $\hat G$ of the measured qubit, are separately multiplied into two distinguished idle and interaction coefficients $A_i$ and $B_i$ of the residual. The proof is provided by Supplementary Note 6.
%
%
\newline
\newline
\textbf{Sign Theorem}: given a register of $N$ qubits, with qubits $i$ and $j$ entangled each other and generated respectively by $\hat G_i(a_i,b_i)$ and $\hat G_j(a_j,b_j)$, after their measurement the residual $\hat R$ displays a term with their interacting coefficients $B_i$, $B_j$, a minus sign in front of it and a set of operators $\hat Z_k$ $\in \mathcal{N}(i,j) = \mathcal{N}(i) \cup \mathcal{N}(j) \setminus \mathcal{N}(i) \cap \mathcal{N}(j) \setminus \mathcal{M}$, where such set denotes the unmeasured qubits previously entangled with either the $i$-th or the $j$-th site:
\begin{equation}
\label{eq:residualCoeffs2}
    - B_i B_j \bigotimes_{k \in \mathcal{N}(i,j)} \hat Z_k
\end{equation}
and $\mathcal{M}$ still denotes the set of the measured qubits. The proof lies in Supplementary Note 7.
\newline
\newline
\textbf{Operator Corollary}: when a single interactive coefficient $i$ holds in a term from the residual, the operators in front of it are set as $\otimes_{k \in \mathcal{N}(i) \setminus {\mathcal{M}}} \hat Z_k$:
\begin{equation}
    A_1 ... A_{i-1} B_i A_{i+1} ... A_n \bigotimes_{k \in \mathcal{N}(i) \setminus \mathcal{M}} \hat Z_k
\end{equation}
The proof of the Corollary is still provided in Supplementary Note 8.
The two Theorems and their Corollary provide a set of rules on how to map a graph into a sequence of operators, a graphic approach quite similar to the Feynman diagrams. Figure \ref{fig:Feynman} portrays a practical example where such rules are applied.
In Supplementary Note 9, a practical application of the two Theorems along with the Corollary is provided. 
\newline
\newline
\textbf{Symmetries on the transfer tensor}
When implementing a single qubit operation, the graph state holds a single output qubit, while the other qubits in the cluster remain to be measured. While the output qubit can be represented via its generator, the measured qubits can be expressed into the residual.
For a single-qubit system, the generator is given by $\hat G_1(a,b) = [a + b \hat X_1]$, where the index $1$ referring to the single qubit can be therefore omitted. In the generator there are no $\hat Z_i$ terms, as no further entangled qubits $i$ are involved. Therefore, following from Equation \eqref{eq:Res1}, in the residual $\hat R $ only a $\hat Z$ operator is present as $\mathcal{N}_i$ consists of the sole output qubit.
The generator and the residual are thus explicitly expressed as
\begin{equation}
\label{eq:unitarySupp1D}
    \hat R \ket{\Psi} = \left[ A + B \hat{Z} \right] \left[ a + b \hat{X} \right] \ket{0}
\end{equation}
where $A$ and $B$ are $\mathbb{C}$-numbers which collect all the coefficients from the residual, while $a$ and $b$ are the coefficients which form the generator $\hat G$ over $\ket{0}$, i.e. the output qubit. From here we introduce the unitary supplement $\hat \eta$, explicitly expressed as
\begin{equation}
    \hat \eta = \begin{bmatrix}
        Aa + Ba & Ab + Bb \\
        Ab - Bb & Aa - Ba
    \end{bmatrix}
\end{equation}
The transformed state $\ket{\psi'}$ is described by the vector supplement $\vec \eta$, consisting of the first column of the unitary supplement:
\begin{equation}
\label{eq:vecsupplement}
    \hat \eta \ket{0} = \vec \eta, \quad 
    \vec \eta = \begin{bmatrix}
        Aa + Ba \\
        Ab - Bb
    \end{bmatrix}
\end{equation}
The corresponding object in the Pauli basis, i.e. the transfer tensor, reads as
\begin{equation}
    \hat T = Aa \hat I + Ab \hat X + Ba \hat Z + Bb \hat Z \hat X = (Aa, Ab, Ba, Bb)
\end{equation}
Even though all of the residuals $\hat R$ from a graph state correspond to a unitary operation -- or at least to a linear combination of $\hat Z_i$ operators -- the reverse condition does not necessary hold. Such hurdle prompts us to study under which condition the MBQC computation can be performed, and eventually leads us to focus on the symmetries on the transfer tensor.
Such symmetry can be expressed as
\begin{equation}
\label{eq:Symm1Index}
    T_0 T_3 = T_1 T_2 \Rightarrow A a B b = A b B a 
\end{equation}
The swap between the $a$, $b$ coefficients can be implemented, in terms of indices, by a finite and discrete group $(i,\circ)$, whose operation $\circ$ implements the transformations in Table \ref{tab:IndexGroup}. Thanks to the action of such group, Equation \eqref{eq:Symm1Index} can be reformulated into the following one:
\begin{equation}
\label{eq:groupSym}
    T_i T_j = T_{i \circ j} T_{j \circ i}
\end{equation}
It is quite straightforward to show that, except for $0\circ 3$ (as $1\circ 2$ as well), all the other labels yield a simple identity, i.e. $T_{i} T_j = T_j T_i$. Such achievement is summarized in Table \ref{tab:IndexGroup}: while almost every couple of elements is closed with respect to the $\circ$ operation, e.g. $i$, $j \to$ $j$, $i$ or the identity $i$, $j\to$ $i$, $j$, the coefficients on the anti-diagonal (in red) overturn such rule.
\begin{table}[h]
\centering
\begin{tabular}{ c || c | c | c | c } 
\, $\circ$ \, & \, $0$ \, & \, $1$ \, & \, $2$ \, & \, $3$ \, \\
\hline
\hline
\, $0$ \, & $0$ & 1 & 0 & \textcolor{red}{1} \\ 
\hline
\, 1 \, & 0 & 1 & \textcolor{red}{0} & 1 \\
\hline 
\, 2 \, & 2 & \textcolor{red}{3} & 2 & 3 \\ 
\hline
3 & \textcolor{red}{2} & 3 & 2 & 3
\end{tabular}
\caption{\textbf{Group operations} $\circ$ \textbf{for the symmetries on transfer tensor T}. How the labels $i$, $j$ of the transfer tensor $\hat T$ transforms under exchange of the $a$, $b$ idle and interactive coefficients, the second one being linked to the $\hat X$ Pauli operator. Such operation is expressed explicitly in Equation \eqref{eq:Symm1Index}. The $i \circ j = k$ operation is summarized in the Table: the first column collects the $i$ terms, the first row the $j$ ones, while the crossed cells the final product $k$. All the operations consist of a swap or an identity operation, with respect to the $i$-th and the $j$-th elements, except for those on the anti-diagonal (in red).}
\label{tab:IndexGroup}
\end{table}
%
When dealing with an $n$-qubits register, the Equation \eqref{eq:groupSym} must be rearranged properly for the $n$-qubit transformations. The residual of the measured qubits, providing a $\hat Z_i$ term for all the output qubits along with the $\hat Z_i \hat Z_j$ combinations:
\begin{equation}
\label{eq:UnitarySuppnqubits}
    \hat \eta_n = \sum_{ m=0 }^n \; \sum_{{\substack{\vec i \subset [1,...,n] \\ \left|i\right|=m}}} \left( C_{\vec i} \bigotimes_{l\in \vec i} \hat Z_{l}\right) \bigotimes_{j=1}^n \hat G(a_j, b_j)
\end{equation}
i.e. the sum runs over all the possible combinations of $\hat Z_l$ operators. For the case $n=2$ in Supplementary Note 10, $\vec i$ varies to be $[0]$, $[1]$, $[2]$ and $[1,2]$, raising the $C_0$, $C_1$, $C_2$ and $C_{12}$ coefficients. The generators are supposed to install non-entangled qubits, therefore $\hat G(a_i,b_i) = a_i + b_i \hat X_i$. The $C_i$ coefficients are $\mathbb{C}$-numbers, which collect all the other coefficients, as from the Sign Theorem and the Operator corollary.
The formalism from Table \ref{tab:IndexGroup} needs to be extended for the general case $n$. Leaving the proof in the Methods, the composition operation translates into the following expression:
\begin{equation}
\label{eq:GroupNQubits}
    T_{\textbf{i}} T_{\textbf{j}} = T_{\textbf{i} \circ \textbf{j}} T_{\textbf{j} \circ \textbf{i}} = T_{(i_1, i_2,..., i_n) \circ (j_1, j_2, ..., j_n)} T_{(j_1, j_2, ..., j_n) \circ (i_1, i_2,..., i_n)}
\end{equation}
which expression involves all the possible compositions between $(i_1, i_2, ..., i_n) \circ (j_1, j_2, ..., j_n)$ and vice versa between $ (j_1, j_2, ..., j_n) \circ (i_1, i_2, ..., i_n)$. A didactic example is provided for the $n=2$ case in the Supplementary Note 10.

\subsubsection*{\label{subsec:MBQCcore}Core equations of MBQC compiling}

A system of equation is eventually formulated, by adding Equation \eqref{eq:GroupNQubits} into the already encoded system of Equations \eqref{eq:FirstSystem}:
%
\begin{subequations}
\label{eq:MBQCompile}
\begin{align}
        T'_{\textbf{i}} = S_{\textbf{ij}} T_{\textbf{j}} = S_{\textbf{ij}} M_{\textbf{jk}} \,\eta_{\textbf{k}},         \label{eq:GraphGen} \\
        \eta_{i0} = \eta'_{i0}, \label{eq:GaugeConstr} \\
        \eta'_{\textbf{i}} = M^{-1}_{\textbf{ij}} T'_{\textbf{j}} = M^{-1}_{\textbf{ij}} S_{\textbf{jk}} T_{\textbf{k}} \label{eq:BackTransf} \\
        T_{\textbf{i}} T_\textbf{j} = T_{\textbf{i} \circ \textbf{j}} T_{\textbf{j} \circ \textbf{i}} \label{eq:TsymmNdim}
\end{align}
\end{subequations}
where $M_{\textbf{i}\textbf{j}}=M_{i_1 j_1}... M_{i_n j_n}$ is the generalized tensor for the change of basis.

\section*{Discussion}

\subsection*{From graphs to the residual and vice versa}


It is therefore possible to introduce a set of techniques to rebuild the graph state from the residual of the unitary operator. In the first place, a decomposition of the unitary matrix $\hat U$ has to be given in terms of Pauli matrices, such as KAK decomposition or any composition of universal gates.
Once such Pauli decomposition is provided, it is possible to transcribe the transfer tensor. However, the resulting outcome may not fit into the relation expressed by Equation \eqref{eq:TsymmNdim}, which requires a gauge transformation to be applied, in order to satisfy both Equations \eqref{eq:TsymmNdim} and \eqref{eq:GaugeConstr}.
Techniques to be applied over a specific gauge transformation, which we call the fully symmetric gauge, are introduced in Supplementary Note 11.
A practical example to exploit the fully symmetric gauge is illustrated in Supplementary Note 12, while an alternative technique is shown in Supplementary Note 13. 
A class of equivalence between graphs corresponding to the same unitary via different gauge transformations is stated in Supplementary Note 14.
Once the new $\hat T'$ transfer tensor is provided, the next step consists of distinguishing between the coefficients belonging to the residual and the coefficients assigned to the generators. To tackle such issue, the Coefficient Theorem turns to be of paramount importance.
The coefficients from the generators report information about the $\hat X_i$ operators, while the coefficients from the residual about the $\hat Z_i$ ones.
From the Sign Theorem, it is possible to check when two interactive coefficients $B_i$, $B_j$ in the transfer tensor output a minus sign, tracking the neighborhood of the measured $i$ and $j$ qubits. Such information reduces by far the number of required operations, as it suffices to check for all the terms involving two $B_i$ coefficients whether a minus sign occurs. Thanks to the Coefficient and the Sign theorems, a set of graphic rules are derived, thanks to which the residual can be mapped into a graph and vice versa. We refer to such rules as \textit{analysis of the residual}, and practical examples are illustrated in Supplementary Notes 9 and 11.
An overall perspective about how to build a graph state from the residual, thanks to the gauge transformation by the tensor $S$, is reported in Algorithm \ref{alg:gauge}.
The fully symmetric gauge proves to reduce the size of the graph state, with respect to the M-Calculus, of $50\%$ when applied for compiling the QFT, and of $75\%$ for the QAOA. The results are shown in Table \ref{Tab:Benchmark}. Further details are shown in Supplementary Note 15.

\begin{table}[h]
\caption{\label{Tab:Benchmark}Number $n$ of involved ancillary qubits in the M-Calculus and fully symmetric gauge. Here $p$ stands for the number of layers set in the QAOA.}\label{tab1}%
\begin{tabular}{@{}lll@{}}
\toprule
                       & M-Calculus  & This work (fully symmetric gauge)\\
\midrule
QFT                    & $3n^2 + 2n$      & $\frac{3}{2} n^2 + \frac{3}{2}n$  \\
QAOA (cyclic Max-Cut)  & $n(1+7p)$        & $2n(1+p)$  \\
QAOA (regular Max-Cut) & $2pn^2 + n(1+p)$ & $\frac{1}{2}pn^2 + n(\frac{3}{2}p - 2 )$  \\
\end{tabular}
\end{table}

\begin{algorithm}[H]
\caption{Gauge Compiling}
\label{alg:gauge}
\begin{algorithmic}[1]
\State Decompose $\hat U$ unitary operator $2^n \times 2^n$ into Pauli basis $\{ \hat I, \hat X_i, \hat Z_i, \hat Z_i \hat X_i \}_{i=1}^n$
\State Build $\hat T$, $\hat \eta$
\State B = False
\While{B}
\State Set a $S_{\textbf{i} \textbf{j}}$ gauge
\State Get $\hat T_{new}$ from Equation \eqref{eq:GraphGen}
\State B = Equations \eqref{eq:GaugeConstr}, \eqref{eq:TsymmNdim} are satisfied
\EndWhile
\State Apply Sign Theorem, Coefficient Theorem on $T$ to get the graph state
\end{algorithmic}
\end{algorithm}

\subsection*{General considerations}

\color{black}
We have introduced a new formalism, based on the generators of a qubit and the residual operator, to describe the graph state and to perform measurement-based computation. Afterwards, we introduced a set of algebraic tools, such as the unitary supplement and the transfer tensor, to define an equivalence class between different graph states implementing the same quantum circuit. The map between the elements of the equivalence class give rise to a gauge transformation between different graph states.
Thereafter, we set the action of the measurement in the new formalism, by defining the residual operator and stating the Coefficient and the Sign Theorems, which allows to reconstruct the graph state starting from the residual -- a technique we call \textit{analysis of the residual}.
In the latter sections, we fix the global picture of this work by writing the core Equations of this scheme of MBQC compiling.
\color{black}

Thanks to such system of equations, different ways to encode a graph state from a single unitary are now available. 
In order to make such Equations work properly, the unitary gate must be decomposed in terms of Pauli gates. Methods such as KAK, or even gate-based circuits provide a suitable decomposition for any unitary operators.
A system of equations turns to be more adaptive than an algorithm, providing specific solutions for any instance of unitary.
Furthermore, deploying the techniques from the statements in the Coefficient Theorem, the Sign Theorem and the following Corollary, the compiling challenge can be tackled from a topological perspective rather than a numerical one, avoiding memory bounds due to the allocation of $n$-qubits matrices with exponential size $2^n$, handling instead a $G[V;E]$ graph.
The actual work provides a useful and universal example of gauge transformation in Supplementary Note 11, nominally the fully symmetric gauge.
Future works may focus on developing numerical and topological techniques to find the most suitable $S$ gauge tensors for a specific unitary to be mapped into a graph state.

\section*{Methods}

\subsection*{Isomorphism between the Pauli and the canonical basis}

As stated from Equation \eqref{eq:Isom}, the $M$ map represents the isomorphism between the canonical $\mathcal{C}$ and the Pauli $\mathcal{P}$ basis:
\begin{subequations}
\label{eq:IsomorphismM}
\begin{align}
    M = \frac{1}{2}
    \begin{pmatrix}
    1 & 0 & 0 & 1 \\
    0 & 1 & 1 & 0 \\
    1 & 0 & 0 & -1 \\
    0 & 1 & -1 & 0
    \end{pmatrix}, \quad M : \mathcal{C} \to \mathcal{P}
    \\
    M^{-1} = \begin{pmatrix}
    1 & 0 & 1 & 0 \\
    0 & 1 & 0 & 1 \\
    0 & 1 & 0 & -1 \\
    1 & 0 & -1& 0 
    \end{pmatrix}, \quad M^{-1} : \mathcal{P} \to \mathcal{C}
\end{align}
\end{subequations}
In a tensor formalism, the $M$ isomorphism acts on each index:
\begin{equation}
    \eta_{\textbf{i}} = \eta_{i_1... i_n} = M_{j_1 k_1} M_{j_2 k_2} ... M_{j_n k_n} T_{k_1 ... k_n} = M_{\textbf{j} \textbf{k}} T_{\textbf{k}}
\end{equation}

\subsection*{Group symmetries on the transfer tensor}
\label{sub:GroupOpN}

In order to generalize to $n$ the group operations from Table from Table \ref{tab:IndexGroup}, we first focus on the $1$-qubits case:
\begin{equation}
\label{eq:Swap1qubit}
    Aa Bb = Ab Ba \Rightarrow T_0 T_3 = T_1 T_2
\end{equation}
We then relabel such coefficients as $C_i$ for the $A$, $B$ and $c_i$ for $a$, $b$, making the previous equation into
\begin{equation}
    T_i T_j = C_i c_i C_j c_j = C_i c_j C_j c_i = T_{i \circ j} T_{j \circ i}
\end{equation}
where $C_0=C_1=A$, $C_2=C_3=B$, $c_0=c_2=a$, $c_1=c_3=b$.
Such method may be adapted for the $n$-qubits system in the following formalism:
\begin{equation}
    T_{\textbf{i}} T_{\textbf{j}} = 
    T_{i_1, i_2, ..., i_n} T_{j_1, j_2, ..., j_n} = C_{i_1, i_2, ..., i_n} c_{i_1} c_{i_2} ... c_{i_n} C_{j_1, j_2, ..., j_n} c_{j_1} c_{j_2} ... c_{j_n}
\end{equation}
In the previous notation, the capital letter $C_{\textbf{i}}$ coefficients refers to the coefficients from the residual in Equation \eqref{eq:UnitarySuppnqubits}, while the lower case $c_{i_k}$ to the idle or active coefficients $a_k$, $b_k$ of the generators $\hat G_k$. As all of the coefficients $c_{i_k}$ of the generators may be exchanged as in Equation \eqref{eq:Swap1qubit}, the composition operation translates into the expression from Equation \eqref{eq:Group2Qubits}:
\begin{equation}
    T_{\textbf{i}} T_{\textbf{j}} = T_{\textbf{i} \circ \textbf{j}} T_{\textbf{j} \circ \textbf{i}} = T_{(i_1, i_2,..., i_n) \circ (j_1, j_2, ..., j_n)} T_{(j_1, j_2, ..., j_n) \circ (i_1, i_2,..., i_n)}
\end{equation}
which expression involves all the possible compositions between $(i_1, i_2, ..., i_n) \circ (j_1, j_2, ..., j_n)$ and vice versa between $(j_1, j_2, ..., j_n) \circ (i_1, i_2, ..., i_n)$.

\bibliography{manuscript}
\endgroup

\section*{Acknowledgements}

S.C. acknowledges Leonardo SPA for having supported his research by a PhD grant.

\section*{Author contributions statement}
Both S.C. and E.P have elaborated and developed the method. They both wrote and reviewed the manuscript. 

\section*{Supporting Information}
Supporting Information is available from the NPG website.

\section*{Data availability}
The data that support the findings of this study are available from the corresponding Author upon reasonable request.

\section*{Accession codes}
No code has been developed in this research.

\section*{Competing Interests statement}
The Authors declare that the patent application 102024000016987 "A computer-implemented method for quantum compiling for measurement-based unidirectional quantum computation, and related system" is pending.

\citestyle{nature}
\bibliographystyle{unsrtnat}


\section*{Supplementary Note 1}
\textbf{Lemma}
The generator operator $\hat G$ is not unitary in general.
\newline
\newline
\textbf{Proof}
The product between a generator $\hat G$ and its adjoint turns to be
\begin{equation}
    \hat G_i^\dagger \hat G_i = \left|a\right|^2 + \left|b\right|^2 + (a^* b + b^* a) \hat K_i = 1 + (a^* b + b^* a) \hat K_i
\end{equation}
Thus the condition for being unitary holds if and only if $(a^* b + b^* a)=0$. $\blacksquare$
\newline
Nevertheless, the expectation of $\hat G_i$ over the zero state $\ket{0}^{\otimes n}$ is equal to one, as the expectation value of $\hat K_i = \hat X_i \bigotimes_{j \in \mathcal{N}(i)} \hat Z_j$ over $\ket{0}^{\otimes n}$ turns to be null:
\begin{equation}
    ^{\otimes n}\braket{0|\hat G_i^\dagger \hat G_i|0}^{\otimes n} = 1
\end{equation}
Such condition is weaker than requiring $\hat G \in U(2^n)$. Nevertheless, if a single qubit $i$ can be described by a generator $\hat G_i$ as in Equation \eqref{eq:EntangledState}, the condition of normalization is still satisfied.

\section*{Supplementary Note 2}

\textbf{Lemma}
If two Pauli operators $\hat K_1$, $\hat K_2$ commute each other with, the new elements $\hat K_1'$, $\hat K_2'$ still commute each other with. In other words, the stabilizer group is closed under the action of the Clifford algebra.
\newline
\newline
\textbf{Proof}
\begin{equation}
    [\hat K_1; \hat K_2] = 0 \Rightarrow 0 = \hat U [\hat K_1; \hat K_2] \hat U^\dagger = [\hat K'_1; \hat K'_2]
\end{equation}
where $\hat U$ is any element from the Clifford algebra. $\blacksquare$
\newline
\newline
\textbf{Corollary}
It is possible to entangle a register of $n$ qubits, represented by an opportune set of $n$ corresponding generators, by applying a set of $\hat{CZ}_{ij}$ gates on the overall state, without affecting the commutativity of the generators.
The new state $\ket{\Psi'}$, in terms of generators, reads now
\begin{equation}
\label{eq:EntangledState}
    \ket{\Psi} = \prod_{i=1}^n [a_i + b_i \hat X_i] \to
    \ket{\Psi'} = \prod_{i=1}^n [a_i + b_i \hat K_i] \ket{0}^{\otimes n} = 
    \prod_{i=1}^n \hat G_i(a_i,b_i) \ket{0}^{\otimes n}
\end{equation}

\section*{Supplementary Note 3}

Let's recall that the overall entanglement can be described by a graph, picturing the qubits as the vertices of such graph and the entanglement as its edges.
The corresponding graph state, composed for instance by two qubits $1$ and $2$, in terms of stabilizers will be simultaneous eigenstate of the $\langle \hat X_1 \hat Z_2; \hat Z_1 \hat X_2 \rangle$ stabilizers. While the $\hat X$ operators denote the nodes of the graph -- or, equivalently, the qubits of the system -- the $\hat Z$ ones refer to its edges, corresponding to the physical entanglement between the qubits~\cite{kruszynska2006entanglement}. Any stabilizer state can be described by a graph, the $\hat X_i$ corresponding to its vertices and the $\hat Z_j$ to the edges. Therefore, the stabilizer states are often referred to as graph states~\cite{benjamin2006brokered,kissinger2019universal}.
The purpose of this Section is to compute the action of the elements of the Clifford group on the generators $\hat G_i$.

The stabilizer group $\mathcal{S}_n$ is defined as the abelian subgroup of the Pauli group $\mathcal{P}_n$:
\begin{equation}
\label{eq:StabGroup}
    \mathcal{S}_n = \{ \hat{S}_i \in \mathcal{P}_n \; | \; \hat{S}_i^\dagger = \hat{S}_i, \; [\hat{S}_i, \hat{S}_j] = 0 \; \forall i,j \}
\end{equation}
A stabilizer state $\ket{\Psi}$ can be described as a simultaneous eigenstate for such a set of commutative operators~\cite{dehaene2003clifford,van2005local,anders2006fast,raussendorf2001one}.
In order to build the entangled states, single and two-qubits unitary operators are required, more specifically the Hadamard $\hat H$ gate and the controlled-phase $\hat{CZ}$ gate. Such operators belong to the so-called Clifford group $\mathcal{C}_n$, which features the property to map any element of the Pauli group into itself~\cite{van2005local,rengaswamy2020logical,kok2009five,browne2016one}:
\begin{equation}
    \mathcal{C}_n = \{ \hat C \in SU(2^n) \; | \; \hat P \in \mathcal{P}_n, \; \hat C^\dagger \hat P \hat C \in \mathcal{P}_n \}
\end{equation}
Some algebraic rules are summed up~\cite{pius2010automatic, danos2007measurement,brell2015generalized}:
\begin{equation}
\label{eq:Clifford}
    \begin{cases}
    \hat H_i^{-1} = \hat H_i \\
    \hat H_i \hat X_i = \hat Z_i \hat H_i \\
    \hat{CZ}_{ij} \hat{X}_i = \hat{X}_i \hat Z_j \; \hat{CZ}_{ij} \\
    \hat{CX}_{ij} = \hat H_j \hat{CZ}_{ij} \hat H_j
    \end{cases}
\end{equation}
$j$ being the target qubit and $i$ the control one.
Therefore, starting with a graph of unconnected vertices and making $\hat{CZ}_{ij}$ promote the connections between them, the overall entanglement of the system can be described by a graph and its adjacency matrix. Conversely, any regular graph can be represented by the stabilizer group, the $\hat X_i$ operators in Equation \eqref{eq:Clifford} pointing to the $i$-th nodes of the graph, the $\hat Z_j$ referring to the $(i,j)$ edges. It is trivial to compute that both the $\hat{CX}$ and $\hat{CZ}$ operators do not affect the $\ket{0}^{\otimes n}$ state.

\section*{Supplementary Note 4}

In literature, the measurements with respect to the Pauli observables can be performed over three planes, the $XY$, the $XZ$ and the $YZ$~\cite{browne2007generalized, joo2019logical,markham2014entanglement}, each of them being a linear combination of two Pauli matrices, whose coefficients are given as $\cos(\theta)$, $\sin(\theta)$. In the generator formalism, such measurements can be respectively expressed as
\begin{eqnarray}
\label{eq:measurements}
    \bra{XY(\theta)} = \bra{0} [\hat I + e^{-i\theta} \hat X] \hat Z^{s}\nonumber\\
    \bra{YZ(\theta)} = \bra{0} \left[\cos(\frac{\theta}{2}) - i \sin(\frac{\theta}{2}) \hat X \right] \hat X^{s}
    \nonumber\\
    \bra{XZ(\theta)} = \bra{0} \left[\cos(\frac{\theta}{2}) + \sin(\frac{\theta}{2}) \hat X \right] \hat X^{s} \hat Z^{s}
\end{eqnarray}
As the outcome from the measurement is probabilistic, in order to make the computation deterministic some corrections need to be fixed. Such adjustment in MBQC literature are called byproducts, and are implemented as Pauli operators $\hat X$, $\hat Z$. A practical way to describe the overall measurement is therefore to exponentiate such byproducts to $s=0,1$, depending on the outcome from the measurement.
In the XY plane, for instance, the observable $\hat O(\theta)$ from which to measure is given by
\begin{equation}
    \hat O(\theta) = \cos(\theta) \hat X + \hat Y \sin(\theta) =
    \begin{pmatrix}
        0 & e^{i\theta} \\
        e^{-i\theta} & 0
    \end{pmatrix}
\end{equation}
The eigenvalues can be evaluated from the characteristic polynomial:
\begin{equation}
    \lambda^2 - 1 = 0 \Rightarrow \lambda_\pm = \pm 1
\end{equation}
Thus, the eigenvalue equation provides the following equations:
\begin{equation}
    \begin{pmatrix}
        0 & e^{i\theta} \\
        e^{-i\theta} & 0
    \end{pmatrix} \begin{pmatrix}
        a \\ b
    \end{pmatrix} =
    \begin{pmatrix}
        b e^{i\theta} \\ a e^{-i\theta}
    \end{pmatrix} 
    \overset{!}{=}
    \pm \begin{pmatrix}
        a \\ b
    \end{pmatrix} 
\end{equation}
whose solutions are given by $b=\pm a e^{-i\theta}$, $\left|a\right|^2 + \left|b\right|^2=1$, and therefore the state vectors are given by
\begin{equation}
    \frac{1}{\sqrt 2}
    \begin{pmatrix}
        1 \\ \pm e^{-i\theta}
    \end{pmatrix} = \frac{1}{\sqrt 2} [\hat I \pm e^{-i\theta} \hat X] \ket{0}
\end{equation}
Therefore, linear combinations of multiple observables allow to introduce rotations, when measuring the graph state. Additionally, it is possible to prove that such rotation on the $XY$ plane is equivalent to a $z$ rotation over the $\ket{+}$ state. Introducing a global phase $e^{i\theta/2}$, the same state vector evolves into
\begin{equation}
    \frac{1}{\sqrt 2} [e^{i\frac{\theta}{2}} \pm e^{-i\frac{\theta}{2}} \hat X] \ket{0} = 
    \frac{1}{\sqrt 2} [\hat R_z(\theta/2) \ket{0} \pm \hat R_z(\theta/2) \ket{1}] 
\end{equation}
which can be easily rearranged into $\hat R_z(\theta/2)\ket{\pm}$.

As quantum measurements are not deterministic, when dealing with qubits in a $\mathbb{C}^2$ Hilbert space, the possible outcomes turn to be $2$. Quantum measurements make the qubit collapse into a bit state, i.e. destroying any superposition, without specifying, however, which bit state they will collapse into. Therefore, when performing MBQC algorithms, a deterministic computation is required. For this reason, some correction operators, called by-products, are introduced whenever the measurements happen. The value of such operators depends on the outcome $s$ of the measurements, as it can be seen from Equations \eqref{eq:measurements}. However, corrections $\hat X_i^s$, $\hat Z_i^s$ must be applied on the unmeasured qubits.

When performing a measurement over the $i$-th qubit on the $YZ$ plane, the transformation on such qubit is implemented as follows:
\begin{equation}
    \bra{0_i} [ C + i S \hat X_i ] \hat X_i^s [ \alpha_i + \beta_i \hat X_i \bigotimes_{k \in \mathcal N(i)} \hat Z_k ] \ket{0_i}
\end{equation}
Erasing the $\hat X_i$ operators, whose expectation value over $\ket{0}$ is null, for $s=0$, $s=1$ the possible results are
\begin{eqnarray}
     C \alpha_i + i S \beta_i \bigotimes_{k \in \mathcal N(i)} \hat Z_k  \nonumber, \qquad
     i S \alpha_i + C \beta_i \bigotimes_{k \in \mathcal N(i)} \hat Z_k
\end{eqnarray}
Thus it follows the statement: if the measured qubit is set on a superposition state, or in any state with $\alpha = e^{i\gamma} \beta$, the byproduct correction consists of applying the $\hat Z$ neighbors of the qubit:
\begin{equation}
     [\bigotimes_{k \in \mathcal N(i)} \hat Z_k]^s  \bra{0_i} [ C + i S \hat X_i ] \hat X_i^s [ \alpha_i + \beta_i \hat X_i \bigotimes_{k \in \mathcal N(i)} \hat Z_k ] \ket{0_i}
\end{equation}

When performing a measurement over the $i$-th qubit on the $XY$ plane, the transformation on such qubit is implemented as follows:
\begin{equation}
    \bra{0_i} [ \hat I + e^{-i \theta} \hat X_i ] \hat Z_i^s [ \alpha_i + \beta_i \hat X_i \bigotimes_{k \in \mathcal N(i)} \hat Z_k ] \ket{0_i}
\end{equation}
Erasing the $\hat X_i$ operators, whose expectation value over $\ket{0}$ is null, for $s=0$, $s=1$ the possible results are
\begin{equation}
     \alpha_i \pm e^{-i \theta} \beta_i \bigotimes_{k \in \mathcal N(i)} \hat Z_k
\end{equation}
If any of the neighbors of the $i$-th qubit is set in a superposition state $\ket{+}$, or rather $\alpha_k = e^{i\gamma} \beta_k$, it is possible to apply the $\hat K_l = \hat X_l \otimes_{j \in \mathcal N(l)} \hat Z_j$ stabilizer, with $l\in \mathcal N(k)$ (i.e. belonging to the neighborhood of $k$) in order to correct the measurement:
\begin{equation}
     [\hat K_l]^s [\alpha_i \pm e^{-i \theta} \beta_i \bigotimes_{k \in \mathcal N(i)} \hat Z_k ] [\alpha_l + e^{i\gamma} \alpha_l \hat K_l]
\end{equation}

\section*{Supplementary Note 5}

\textbf{Proof}
To compute how a generic measurement $\bra{\gamma_i}$ affects the overall state in the formalism of the generators, first commute the $i$-th generator up to the leftmost position. Such operation is always allowed thanks to the commutativity of the stabilizer group:
\begin{equation}
    \ket{\Psi} = [a_i + b_i \hat K_i] \bigotimes_{\substack{j=1 \\ j\neq i}}^n \hat G_j(a_j, b_j) \ket{0}^{\otimes n}
\end{equation}
Thereafter, contract the $\ket{0_i}$ state to the $i$-th generator, which has been granted to be possible by the previous lemma.
\begin{equation}
    [a_i + b_i \hat K_i] \ket{0_i} \bigotimes_{\substack{j=1 \\ j\neq i}}^n \hat G'_j(a_j, b_j) \ket{0}^{\otimes n-1}
\end{equation}
where $G'$ are the generators affected by such operation, as described in the previous lemma.
In second place, apply the $\bra{\gamma_i} = \bra{0_i} [ \alpha_i^* + \beta_i^* \hat{X}_i]$ measurement over the $i$-th generator.
The residual is yielded by the following expectation value over the $i$-th qubit:
\begin{eqnarray}
    \hat R = \bra{0_i} [ \alpha_i^* + \beta_i^* \hat{X}_i] [a_i + b_i \hat{X}_i \bigotimes_{j \in \mathcal{N}(i)} \hat Z_j] \ket{0_i} = \nonumber\\ \alpha_i^* a_i + \beta_i^* b_i \bigotimes_{j \in \mathcal{N}(i)} \hat Z_j = A_i + B_i \bigotimes_{j \in \mathcal{N}(i)} \hat Z_j
\end{eqnarray}
which expression matches Equation \eqref{eq:Res1}.
$\mathcal{N}(i)$, again, being the set of qubits entangled with the $i$-th qubit before its measurement, and $A_i = \alpha_i^* a_i$, $B_i=\beta_i^* b_i$.
No $\hat X_i$ terms appear anymore, as $\bra{0} \hat X \ket{0}=0$. The residual operator $\hat R$ can be settled as the operator which acts on the unmeasured qubits, as a consequence of the entanglement with the previously measured ones. The new $\ket{\Psi'}$ state, after such measurement, now reads as
\begin{equation}
    \ket{\Psi'} = \hat R \; \bigotimes_{\substack{j=1 \\ j\neq i}}^n \hat G'(a_j, b_j) \ket{0}^{\otimes n-1}
\end{equation}
To complete the proof, one recursively applies the measurement operation $\bra{\gamma_l}$ over the $\ket{\Psi'}$ state, as follows:
\begin{equation}
\begin{split}
    \bra{\gamma_l} \ket{\Psi'} = \bra{0_l} [\alpha_l^* + \beta^*_l \hat X_l] \hat R [a_l + b_l \hat K_l] \ket{0_l} \\
    \bigotimes_{\substack{j=1 \\ j\neq i, l}}^n \hat G''(a_j, b_j) \ket{0}^{\otimes n-2}
\end{split}
\end{equation}
As $\hat K_l = \hat X_l \bigotimes_{f \in \mathcal{N}(l)} \hat Z_f$ and $\hat R = A_i + B_i \bigotimes_{j \in \mathcal{N}(i)} \hat Z_j$, the new residual reads
\begin{equation}
\label{eq:Rrecursion0}
    \alpha_l^* \hat R' a_l + \beta^*_l b_l \bra{0_l} \hat X_l \hat R \hat K_l] \ket{0_l} 
\end{equation}
where $\hat R' = A_i + B_i \bigotimes_{j \in \mathcal{N}(i) \setminus \{l\}} \hat Z_j$, as $\bra{0_l} \hat Z_l \ket{0_l} = 1$. From the new operator $\bra{0_l} \hat X_l \hat R \hat K_l] \ket{0_l}$ no $\hat X_l$ term survives, therefore the new residual is still not composed by any $\hat X$ operator, like it occurred during the first iteration in Equation \eqref{eq:Res1}. It is possible to relabel the new residual as $\hat R$
\begin{equation}
\label{eq:Rrecursion}
    \hat R \leftarrow \alpha_l^* \hat R' a_l + \beta^*_l b_l \bra{0_l} \hat X_l \hat R \hat K_l] \ket{0_l}
\end{equation}
and iterate such procedure recursively for any measurement. $\blacksquare$

\section*{\label{Note:Coeff}Supplementary Note 6}

\textbf{Proof (Coefficient Theorem)}: recalling Equation \eqref{eq:Rrecursion}, it follows that
\begin{equation}
    \hat R \leftarrow A_i \hat R' + B_i \bra{0_l} \hat X_l \hat R \hat K_l] \ket{0_l}
\end{equation}
As such operation is repeated recursively, it holds for any measurement. $\blacksquare$

\section*{Supplementary Note 7}

\textbf{Proof (Sign Theorem)}: it follows from Equations \eqref{eq:Rrecursion0} and \eqref{eq:Rrecursion}. The second term from Equation \eqref{eq:Rrecursion0} can be rearranged as
\begin{equation}
\begin{split}
 B_l \bra{0_l} \hat X_l \hat R \hat K_l] \ket{0_l} =
    A_i B_l \bra{0_l} \hat X_l \hat K_l \ket{0_l} + B_i B_l \bra{0_l} \hat X_l \bigotimes_{j \in \mathcal{N}(i)} \hat Z_j \hat K_l \ket{0_l}
\end{split}
\end{equation}
The $A_i B_l$ belongs to the first order of recursion, as in Equation \eqref{eq:Res1}:
%
\begin{equation}
\label{eq:FirstOrder}
    A_i B_l \bra{0_l} \hat X_l \hat K_l \ket{0_l} = A_i B_l \bigotimes_{w \in \mathcal{N}(l)} \hat Z_w
\end{equation}
while the second term reads
\begin{equation}
\begin{split}
    B_i B_l \bra{0_l} \hat X_l \bigotimes_{j \in \mathcal{N}(i)} \hat Z_j \hat K_l \ket{0_l} =
    (-)^s B_i B_l \bra{0_l} \bigotimes_{j \in \mathcal{N}(i)} \hat Z_j \hat X_l \hat K_l \ket{0_l} =
    (-)^s B_i B_l \bigotimes_{j \in \mathcal{N}(i)\setminus \{l\}} \hat Z_j \bigotimes_{w \in \mathcal{N}(l)} \hat Z_w
\end{split}
\end{equation}
where $s\in\{0,1\}$ represents the condition $l \in \mathcal{N}(i)$. If it holds, then the $\hat X_l \hat Z_l$ operators anti-commute, yielding a minus sign and thus $s=1$, otherwise, if no $\hat Z_l$ is present (because $l \notin \mathcal{N}(i)$), then $s=0$. The final expression reads
\begin{equation}
\begin{split}
    B_l \bra{0_l} \hat X_l \hat R \hat K_l] \ket{0_l} =
    A_i B_l \bigotimes_{w \in \mathcal{N}(l)} \hat Z_w + (-)^s B_i B_l \bigotimes_{j \in \mathcal{N}(i)\setminus \{l\}} \hat Z_j \bigotimes_{w \in \mathcal{N}(l)} \hat Z_w
\end{split}
\end{equation}
The two tensor products of the right term can be decomposed as follows:
\begin{equation}
    \bigotimes_{j \in \mathcal{N}(i) \cap \mathcal{N}(l) \setminus \{l\}} \hat Z_j^2 \bigotimes_{w \in \mathcal{N}(l) \cup \mathcal{N}(i) \setminus \mathcal{N}(l) \cap \mathcal{N}(i) \setminus \{l\}} \hat Z_w
\end{equation}
from which the assumption of the theorem, setting $\{l\} = \mathcal{M}$. A practical example is provided in Figure (\ref{fig:neighborhood}). As such procedure can be iterated, it holds for any measurement.
%
$\blacksquare$

\section*{Supplementary Note 8}

\textbf{Proof (Operator Corollary)}: As the idle coefficients are not associated with any operator, recalling Equation \eqref{eq:FirstOrder} the interactive terms $B_i$ in the residual display a $\hat Z_k$ for the output qubits $k \in \mathcal{N}(i)$:
\begin{equation}
    A_1 ... A_{i-1} B_i A_{i+1} ... A_n \bigotimes_{k \in \mathcal{N}(i)} \hat Z_k
\end{equation}
$B_i$ being the sole interactive coefficient in the sequence of $A_j$, $j \neq i$. If more interactive terms would contribute to the sequence, the Sign Theorem should be invoked.

\section*{Supplementary Note 9}

\begin{figure}[ht]
\includegraphics[width=6cm]{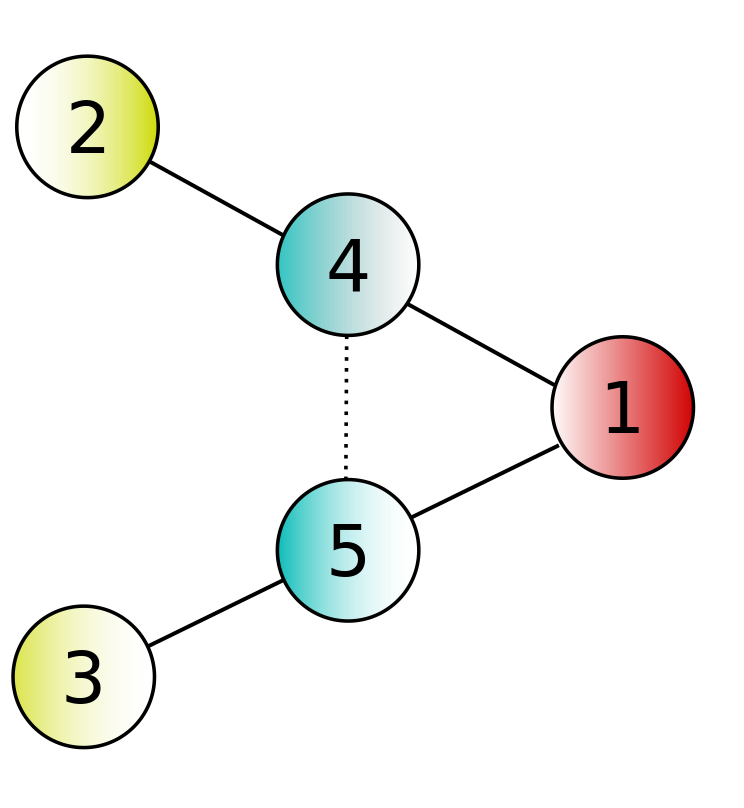}
\caption{Two qubits $4$ and $5$ (in light blue) are measured. The yellow qubits are entangled with $4$ or $5$, while the red one with both of them. After the measurements, the generator of $1$ switches from $\hat X_1 \hat Z_4 \hat Z_5 \to \hat X_1$ and so on for $2$ and $3$.
The interactive coefficients of $4$ and $5$ in the residual, therefore, yield the $(-)^sB_4 \hat Z_1 \hat Z_2 \, B_5 \hat Z_1 \hat Z_3$ term. As both the qubits are entangled with the $1$-th one, $1 \in \mathcal{N}(4) \cap \mathcal{N}(5)$, and therefore no $\hat Z_1$ operator appears. If the dashed line were a solid one, i.e. the two qubits were entangled, it holds $s=1$, otherwise $s=0$ and no line appears between $a$ and $b$. The qubits $2$ and $3$ (yellow) belong to $N(4,5)=N(4)\cup N(5) \setminus \mathcal N(4) \cap N(5)$, while the $1$-st one (red) to $\mathcal N(4) \cap \mathcal N(5)$, thus it is erased from the $B_4 B_5$ term.}
\label{fig:neighborhood}
\end{figure}

In this Section, we deploy the case of a $5$-qubits graph state built as in Figure \ref{fig:neighborhood} where $4$ and $5$ qubits are measured, resulting in a $3$-qubits output state. Such process provides a practical example to apply the two theorems till now exposed. In order to assess how the entanglement between measured qubits affects the minus sign as stated by the Sign Theorem, we compare two cases where the $4$ and $5$ qubits may be entangled or not. In the first place, we write down the generators of the system, following the rules of entanglement as expressed in Equation \eqref{eq:EntangledState}:
\begin{equation}
    \ket{\Psi} = [a_1 + b_1 \hat X_1 \hat Z_4 \hat Z_5] [a_2 + b_2 \hat X_2 \hat Z_4] [a_3 + b_3 \hat X_3 \hat Z_5]
    [a_4 + b_4 \hat X_4 \hat Z_2 \hat Z_1 \hat Z_5^s] [a_5 + b_5 \hat X_5 \hat Z_1 \hat Z_3 \hat Z_4^s] \ket{0}^{\otimes 5}
\end{equation}
$s \in \{0, 1\}$ depending whether the entanglement occurs between $a$ and $b$, represented by the dashed line in Figure \ref{fig:neighborhood}. We then choose to measure the qubit $a$, which measurement sets the overall state $\ket{\Psi_5} \to \ket{\Psi_4}$:
\begin{equation}
    \braket{\gamma_4|\Psi_5} = \bra{0_4} [\alpha_4 + \beta_4 \hat X_4] [a_4 + b_4 \hat X_4 \hat Z_2 \hat Z_1 \hat Z_5^s] \ket{0_4} 
    [a_1 + b_1 \hat X_1 \hat Z_5] [a_2 + b_2 \hat X_2] [a_3 + b_3 \hat X_3 \hat Z_5] 
    [a_5 + b_5 \hat X_5 \hat Z_1 \hat Z_3] \ket{0}^{\otimes 4} = \hat R \ket{\Psi_4}
\end{equation}
The expectation over $\ket{0_4}$ yields the residual of the measurement:
\begin{equation}
    \hat R = \bra{0_4} [\alpha_4 + \beta_4 \hat X_4] [a_4 + b_4 \hat X_4 \hat Z_1 \hat Z_2 \hat Z_5^s] \ket{0_4} = 
    \alpha_4 a_4 + \beta_4 b_4 \hat Z_1 \hat Z_2 \hat Z_5^s = 
    A_4 + B_4 \hat Z_1 \hat Z_2 \hat Z_5^s
\end{equation}
Afterwards, we perform the measurement over the qubit $5$ on the $\ket{\Psi_4}$ state:
\begin{equation}
    \bra{\gamma_5} \hat R \ket{\Psi_4} = \bra{0_5} [\alpha_5 + \beta_5 \hat X_5] \hat R [a_5 + b_5 \hat X_5 \hat Z_1 \hat Z_3] \ket{0_5}
    [a_1 + b_1 \hat X_1] [a_2 + b_2 \hat X_2] [a_3 + b_3 \hat X_3] \ket{0}^{\otimes 3}
\end{equation}
We now compute the residual after the second measurement:
\begin{equation}
    \bra{0_5} [\alpha_5 + \beta_5 \hat X_5]  [A_4 + B_4 \hat Z_1 \hat Z_2 \hat Z_5^s]  [a_5 + b_5 \hat X_5 \hat Z_1 \hat Z_3] \ket{0_5} =
    A_4 A_5 + A_4 B_5 \hat Z_1 \hat Z_3 + B_4 A_4 \hat Z_1 \hat Z_2 +
    B_4 B_5 \bra{0_5} \hat X_5 \hat Z_2 \hat Z_3 \hat Z_5^s \hat X_5 \ket{0_5}
\end{equation}
Depending on $s=0$ or $s=1$, it holds a minus sign in front of the two interactive coefficients $B_4 B_5$. After the two measurements, the residual reads
\begin{equation}
\label{eq:ExampleFig}
    \hat R =
    A_4 A_5 + A_4 B_5 \hat Z_1 \hat Z_3 + B_4 A_5 \hat Z_1 \hat Z_2
    + (-)^s B_4 B_5 \hat Z_2 \hat Z_3
\end{equation}
As stated by the Corollary Operator, each interactive coefficient carries a sequence of $\hat Z_i$ operators, $i$ pointing to the output qubits the measured ones were entangled with. The Sign Theorem, instead, states that if two measured qubits were entangled, a minus sign arises in front of their interactive coefficients. Thanks to such theorems, it is possible to achieve the result in Equation \eqref{eq:ExampleFig} avoiding all of the algebraic calculations.

\section*{Supplementary Note 10}

In this Section, we adapt the multi-qubits formalism for the gauge transformations to the particular case $n=2$, in order to ease the comprehension of the tensor depiction of the MBQC compiling equations.
In order to introduce transformations on systems of two qubits, we describe at first the unitary supplement for such system, assuming the output qubits not to be entangled:
\begin{equation}
    \label{eq:residual2D}
    \hat \eta_2 = \left[ C_0 + C_1 \hat Z_1 + C_2 \hat Z_2 + C_{12} \hat Z_1 \hat Z_2 \right] \bigotimes_{j=1}^2 \hat G(a_j, b_j)
\end{equation}
The next step consists of representing the transfer tensor $T$.
%
%
From a $1$-qubit system to a $2$-qubits one, $\hat T$ scales from a $1$-rank $T_i$ tensor to a $2$-rank $T_{ij}$. Its expression is therefore reshaped into
\begin{equation}
\label{eq:2qubitsT}
    \begin{bmatrix}
        C_0 a_1 a_2 & C_0 a_1 b_2 &  C_2 a_1 a_2 &  C_2 a_1 b_2 \\
        C_0 b_1 a_2 & C_0 b_1 b_2 &  C_2 b_1 a_2 &  C_2 b_1 b_2 \\
        C_1 a_1 a_2 & C_1 a_1 b_2 &  C_{12} a_1 a_2 &  C_{12} a_1 b_2 \\
        C_1 b_1 a_2 & C_1 b_1 b_2 &  C_{12} b_1 a_2 &  C_{12} b_1 b_2 \\
    \end{bmatrix}
\end{equation}
%
The table reporting all the coefficients with respect to their generators in the Pauli basis is shown in Table \ref{tab:PauliTransfer2Dtensor}. 
The transformations induced by $M$ and $M^{-1}$ can be generalized to
\begin{equation}
    T'_{i_1 j_1} = S_{i_1 i_2 j_1 j_2} T_{i_2 j_2} =  S_{i_1 i_2 j_1 j_2} M_{i_2 i_3} M_{j_2 j_3} T_{i_3 j_3}
\end{equation}
The following compact notation on the indices helps lightening such formalism:
\begin{equation}
\label{eq:2indexCore}
    T'_{\textbf{i}} = S_{\textbf{i} \textbf{j}} T_{\textbf{j}} =  S_{\textbf{i} \textbf{j}} M_{\textbf{j}\textbf{k}} T_{\textbf{k}}
\end{equation}
In such case, $M_{\textbf{i} \textbf{j}} = M_{i_2 i_3} M_{j_2 j_3}$, while the gauge transformation $S_{\textbf{i} \textbf{j}}$ is not constrained to factorize into two $2$-rank tensors.

\begin{table}
\centering
\vspace{.2cm}
\begin{tabular}{ |c|c|c|c| }
\hline
$I$ & $X_2$ & $Z_2$ & $Z_2 X_2$ \\
\hline
$X_1$ & $ X_1 X_2$ & $ X_1  Z_2$ & $ X_1  Z_2  X_2$ \\ 
\hline
$Z_1$ & $ Z_1 X_2$ & $ Z_1 Z_2$ & $Z_1 Z_2 X_2$ \\
\hline 
$Z_1 X_1$ & $Z_1 X_1 X_2$ & $Z_1 X_1 Z_2$ & $Z_1 X_1 Z_2 X_2$ \\ 
\hline
\end{tabular}
\caption{The Table shows how to arrange, in the tensor format, the coefficients to their respective generators from the Pauli basis, which are described as linear combination in Equation \eqref{eq:residual2D}. The transfer tensor for th single-qubit case, indeed, matches the first row (or first column) of the above table.}
\label{tab:PauliTransfer2Dtensor}
\end{table}
The group operations for the symmetries over the transfer tensor $T$ need to be explained for the case $n=2$. In the first place, we may review the group operations as swaps between the coefficients of the generator:
\begin{equation}
    Aa Bb = Ab Ba \Rightarrow T_0 T_3 = T_1 T_2
\end{equation}
We may relabel such coefficients as $C_i$ for the $A$, $B$ and $c_i$ for $a$, $b$, making the previous equation into
\begin{equation}
    T_i T_j = C_i c_i C_j c_j = C_i c_j C_j c_i = T_{i \circ j} T_{j \circ i}
\end{equation}
where $C_0=C_1=A$, $C_2=C_3=B$, $c_0=c_2=a$, $c_1=c_3=b$.
Such method may be adapted for the 2-qubits system in the following fashion:
\begin{equation}
    T_{i_1, j_1} = C_{i_1, j_1} c_{i_1} c_{j_1}
\end{equation}
where, referring to the tensor format in Equation \eqref{eq:2qubitsT}, $C_{0,0} = C_{0,1} = C_{1,0} = C_{1,1} = C_0$, then $C_{2,0} = C_{2,1} = C_{3,0} = C_{3,1} = C_1$ and so on. Indeed, the above expression can be read as an identity:
\begin{equation}
    T_{i_1, j_1} = C_{i_1, j_1} c_{i_1} c_{j_1} = T_{i_1 \circ i_1, j_1 \circ j_1}
\end{equation}
which composition returns, in fact, the identity. Such consideration nonetheless leads to a non trivial result:
\begin{equation}
    T_{i_1, j_1} T_{i_2, j_2} = C_{i_1, j_1} c_{i_1} c_{j_1} C_{i_2, j_2} c_{i_2} c_{j_2}
    = C_{i_1, j_1} c_{i_2} c_{j_1} C_{i_2, j_2} c_{i_1} c_{j_2} = T_{i_1 \circ i_2, j_1} T_{i_2 \circ i_1, j_2}
\end{equation}
and the same operation may be repeated for $j_1$ and $j_2$. Group compositions between different indices (e.g. $i_k$ and $j_k$) commute each other.
Such freedom can be encoded in the following formalism:
\begin{equation}
\label{eq:Group2Qubits}
    T_{\textbf{i}} T_{\textbf{j}} = T_{\textbf{i} \circ \textbf{j}} T_{\textbf{j} \circ \textbf{i}} = T_{(i_1, j_1) \circ (i_2, j_2)} T_{(i_2, j_2)\circ (i_1, i_1)}
\end{equation}
which expression involves all the possible compositions between $(i_1, j_1) \circ (i_2, j_2)$ and vice versa, including the trivial ones (the identity):
\begin{equation}
    \begin{cases}
        T_{i_1 \circ i_1, j_1 \circ j_1} T_{i_2 \circ i_2, j_2 \circ j_2} \\
        T_{i_1 \circ i_2, j_1} T_{i_2 \circ i_1, j_2} \\
        T_{i_1, j_1 \circ j_2} T_{i_2, j_2 \circ j_1} \\
        T_{i_1 \circ i_2, j_1 \circ j_2} T_{i_2 \circ i_1, j_2 \circ j_1}
    \end{cases}
\end{equation}

\section*{Supplementary Note 11}

In this Section, we introduce the fully symmetric gauge $S^f$, a universal gauge which fits for any unitary transformation. For the single-qubit case, the gauge operator reads as
\begin{equation}
    S^{f} = \frac{1}{2} \begin{pmatrix}
        1 & 1 & 1 & -1 \\
        1 & 1 & 1 & -1 \\
        1 & -1 & 1 & 1 \\
        1 & -1 & 1 & 1 \\
    \end{pmatrix}
\end{equation}
It is straightforward to compute $\det(S^f)=0$, which makes such transformation a singular one. We now briefly prove how such $1d$ transformation can be achieved.
The action of the unitary supplement, for the $n=1$ qubit case, can be explicitly written as
\begin{equation}
    \hat{\eta} \ket{0} = \begin{pmatrix}
    \eta_{00} & \eta_{01} \\
    \eta_{10} & \eta_{11}
    \end{pmatrix} \begin{pmatrix}
    1 \\ 0
    \end{pmatrix}
\end{equation}
Recall the isomorphism $M$ in Equation \eqref{eq:IsomorphismM} between the canonical and the Pauli bases:
\begin{equation}
    M = \frac{1}{2}
    \begin{pmatrix}
    1 & 0 & 0 & 1 \\
    0 & 1 & 1 & 0 \\
    1 & 0 & 0 & -1 \\
    0 & 1 & -1 & 0
    \end{pmatrix}, \quad M : \mathcal{C} \to \mathcal{P}, \qquad
    M^{-1} = \begin{pmatrix}
    1 & 0 & 1 & 0 \\
    0 & 1 & 0 & 1 \\
    0 & 1 & 0 & -1 \\
    1 & 0 & -1& 0 
    \end{pmatrix}, \quad M^{-1} : \mathcal{P} \to \mathcal{C}
\end{equation}
It is possible to set the following choice of gauge when building the transfer matrix, and thus the graph state:
\begin{equation}
\label{eq:Gauge}
    \begin{cases}
        T_{00} \to T_{00} + \gamma = \Tilde{T}_{00} \\
        T_{01} \to T_{01} + \gamma' = \Tilde{T}_{01} \\
        T_{10} \to T_{10} - \gamma = \Tilde{T}_{10} \\
        T_{11} \to T_{11} + \gamma' = \Tilde{T}_{11}
    \end{cases}
\end{equation}
with $\gamma$, $\gamma'$ being two constants, and $\Tilde{T}$ the new transfer matrix. In fact, such operation makes the $\vec{\eta}$ invariant, according to Equations \eqref{eq:IsomorphismM} in fact $\eta_{00} = T_{00}+\gamma + T_{10} - \gamma$ and $\eta_{10} = T_{01} + \gamma' - (T_{11} + \gamma')$. Thus, the relation $\vec \eta' = \vec \eta$ is satisfied. Imposing the symmetries on the new transfer tensor $\Tilde{T}$, it is possible to get the values for $\gamma$ and $\gamma'$:
\begin{equation}
    \begin{cases}
    {T}'_{00} = {T}'_{01} \\
    {T}'_{10} = {T}'_{11}
    \end{cases}
    \Rightarrow
    \begin{cases}
    T_{00} + \gamma = T_{01} + \gamma' \\
    T_{10} - \gamma = T_{11} + \gamma'
    \end{cases}
\end{equation}
which system always admits a solution:
\begin{equation}
    \begin{cases}
    \gamma' = \frac{1}{2}(T_{00} + T_{10} - T_{01} - T_{11}) \\
    \gamma  = \frac{1}{2}(T_{01} + T_{10} - T_{00} - T_{11})
    \end{cases}
\end{equation}
Of course, when the transfer matrix is already symmetric, the $\gamma$ and $\gamma'$ vanish, as far as $T_{00}=T_{01}$ and $T_{10}=T_{11}$. The new transfer matrix $\Tilde{T}$ thus reads as
\begin{equation}
\label{eq:Tsymm}
    {T}' = \frac{1}{2} \begin{bmatrix}
    T_{00} + T_{01} + T_{10} - T_{11}  &
    T_{00} + T_{01} + T_{10} - T_{11} \\
    T_{00} - T_{01} + T_{10} + T_{11}  &
    T_{00} - T_{01} + T_{10} + T_{11}
    \end{bmatrix}
\end{equation}
$T_{ij}$ being the coefficient of the original transfer matrix $T$. It is quite straightforward to show that $T'$ now satisfied the constraint of symmetry $T_{i} T_j = T_{i\circ j} T_{j \circ i}$.
The same transformation can be better represented in the tensor format:
\begin{subequations}
\begin{align}
    T'_i = S_{ij} T_j  \label{subeq:TgaugesymmApp}\\
    \begin{pmatrix}
        T'_{00} \\ T'_{01} \\ T'_{10} \\ T'_{11}
    \end{pmatrix} = 
    \frac{1}{2}
    \begin{pmatrix}
        1 & 1 & 1 & -1 \\
        1 & 1 & 1 & -1 \\
        1 & -1 & 1 & 1 \\
        1 & -1 & 1 & 1 \\
    \end{pmatrix}
    \begin{pmatrix}
        T_{00} \\ T_{01} \\ T_{10} \\ T_{11}
    \end{pmatrix}
\end{align}
\end{subequations}
The Equation \eqref{subeq:TgaugesymmApp} indeed matches the cardinal Equation for single qubit system
\begin{equation}
    T'_i = S_{ij} T_j = S_{ij} M_{jk} \eta_k
\end{equation}
The same transformation for the 1-qubit case can be described via the cardinal equation
\begin{equation}
    \eta'_i = M_{ij} T_{j} = M_{ij} S_{jk} T_{k}
\end{equation}
Here $\eta_{0} = \eta_{00}$ and $\eta_2 = \eta_{10}$, which are the elements belonging to the vector supplement we want to preserve. Indeed, in order to check this fact, a manipulation of the previous equation turns to be useful:
\begin{equation}
    \eta'_i = S'_{ik} T_{k}
\end{equation}
where $S'_{ik} = M_{ij} S_{ij}$. Therefore, in order to preserve $\vec \eta$, we want $S'_{0j} = M_{0j}$ and $S'_{2j} = M_{2j}$, so that the elements from the vector supplement are still preserved. By a simple computation, for the fully symmetric gauge $S^f$ we achieve
\begin{equation}
    M_{ij} S^f_{jk} = \begin{pmatrix}
        1 & 0 & 1 & 0 \\
        0 & 1 & 0 & 1 \\
        0 & 1 & 0 & -1 \\
        1 & 0 & -1 & 0 \\
    \end{pmatrix}
    \frac{1}{2}
    \begin{pmatrix}
        1 & 1 & 1 & -1 \\
        1 & 1 & 1 & -1 \\
        1 & -1 & 1 & 1 \\
        1 & -1 & 1 & 1 \\
    \end{pmatrix} = 
    \begin{pmatrix}
        1 & 0 & 1 & 0 \\
        1 & 0 & 1 & 0 \\
        0 & 1 & 0 & -1 \\
        0 & 1 & 0 & -1
    \end{pmatrix}
\end{equation}
Hence, $S'_{0j} = (1, 0, 1, 0) = M_{0j}$, $S'_{2j} = (0, 1, 0, -1) = M_{2j}$, which accomplish our purpose. The multi-qubit case is quite straightforward, as the fully symmetric gauge $S^f_n$ is introduced as a factorized product of the single-qubit version $S^f$:
\begin{equation}
    S^f_n = \bigotimes_{i=1}^n S^f
\end{equation}
Therefore, the previous cardinal equation can be rewritten as
\begin{equation}
    \eta_{i_1...i_n} = S^{f}_{i_1... i_n j_1...j_n} M_{j_1 k_1} ... M_{j_n k_n} T_{k_1 ... k_n} = S^{f}_{i_1 j_1} ... S^f_{ i_n j_n} M_{j_1 k_1} ... M_{j_n k_n} T_{k_1 ... k_n} = S^{'f}_{i_1 k_1} ... S^{'f}_{ i_n k_n} T_{k_1 ... k_n}
\end{equation}
Nonetheless, each of the $S'f_{i_l k_l}$ tensors obeys to the constraint $S'f_{0 j} = M_{0j}$, proving the scalability of the $S^f_n$ gauge along with $n$.

\begin{figure}
\subfloat[\label{subfig:FullSymm}]{{\includegraphics[width=0.3\textwidth]{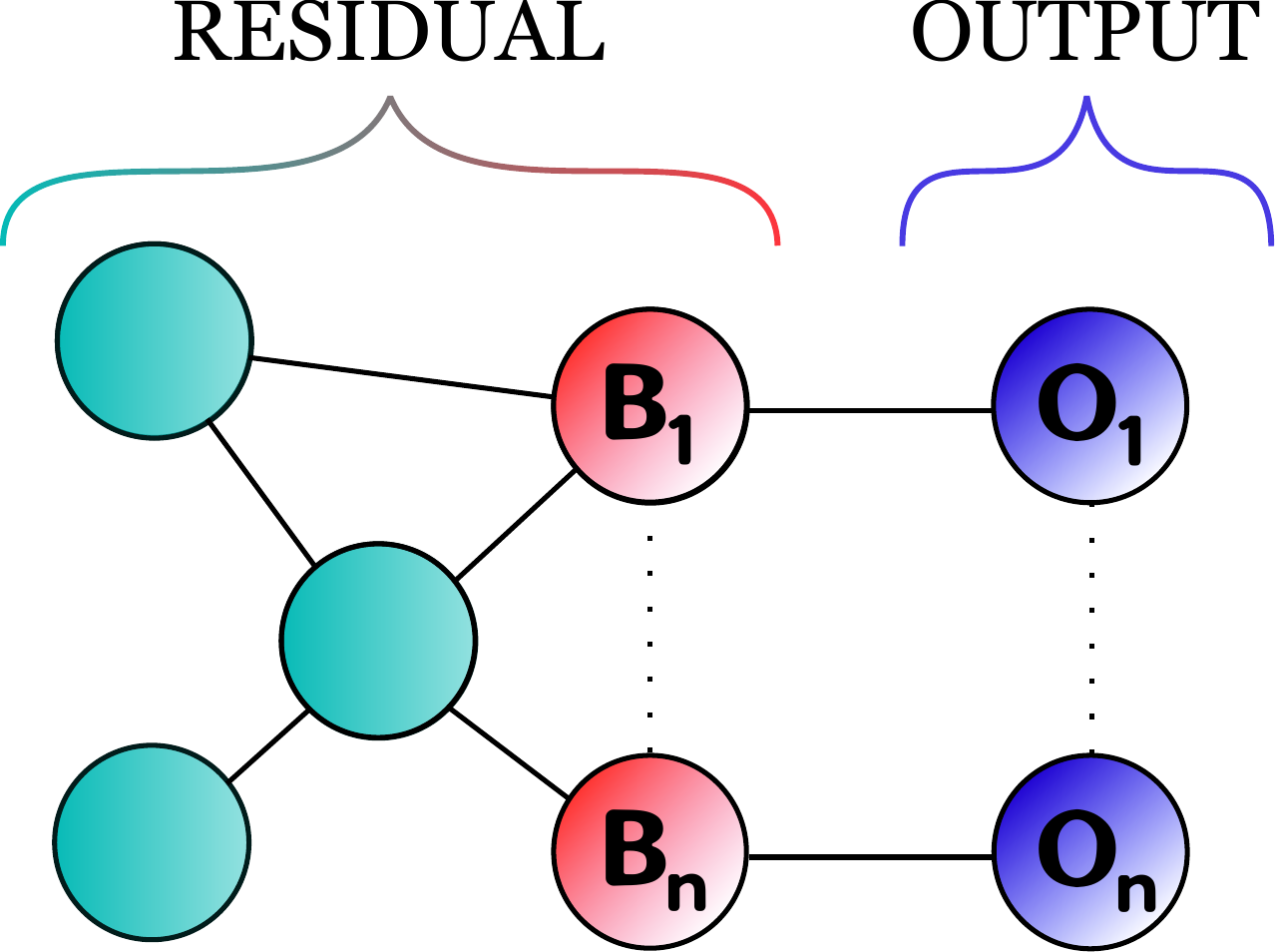} }}
\quad
\subfloat[\label{subfig:Rx}]{{\includegraphics[width=0.3\textwidth]{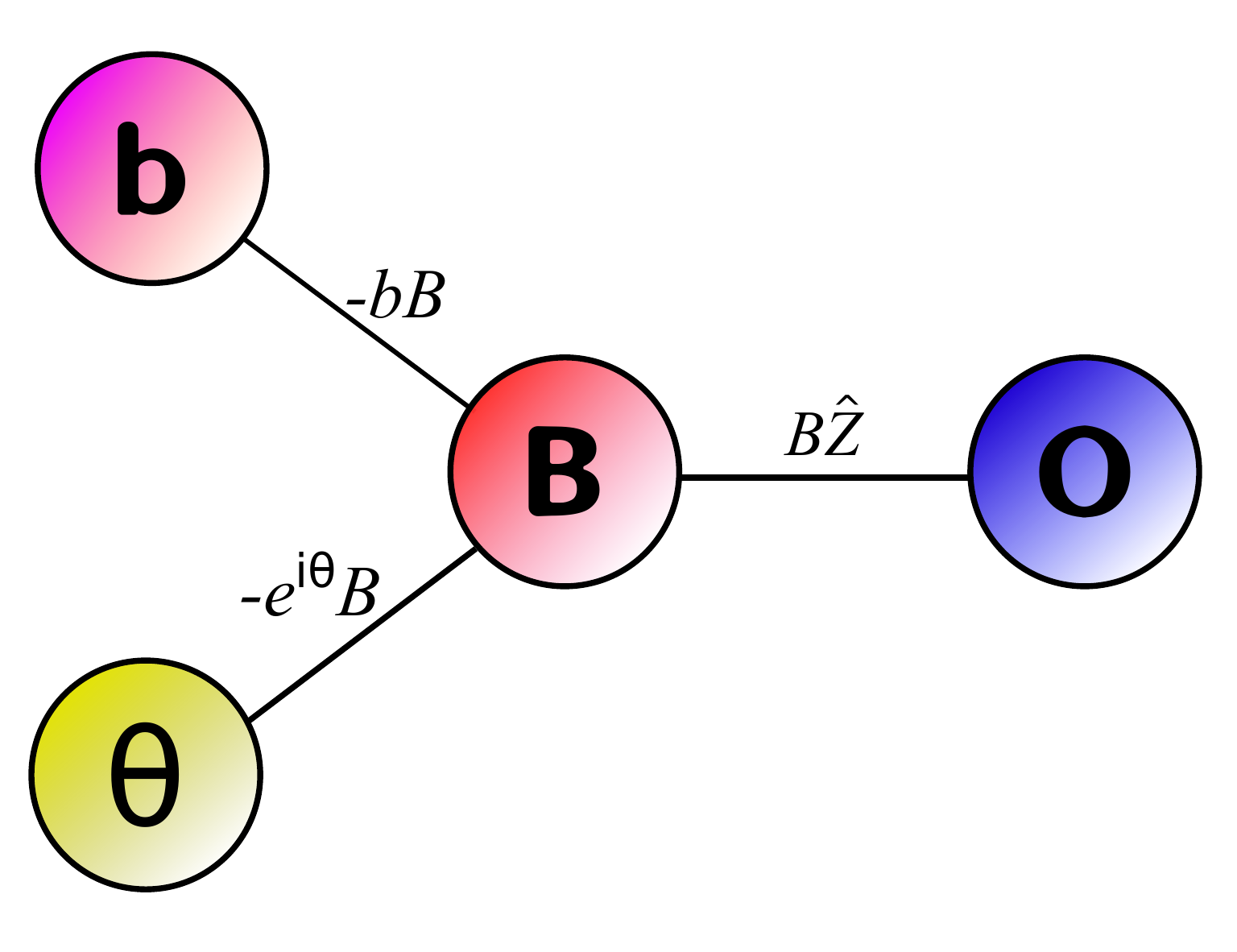} }}
\quad
\subfloat[\label{subfig:Rx1}]{{\includegraphics[width=0.3\textwidth]{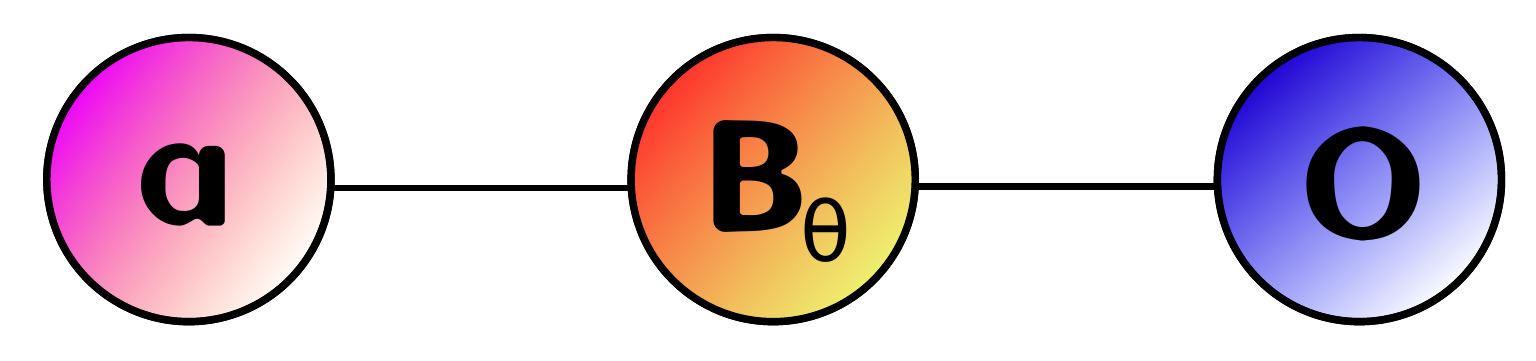} }}
\caption{(a) The fully symmetric gauge outputs a graph state where the outputs are set in a $\ket{+}$ state, linked to a second layer of qubits in the $\ket{+}$ state. From such layer, all the connections with the residual are established. (b) The graph state for a $R_x(\theta)$ rotation resulting from a fully symmetric gauge, as exposed in Supplementary Note 12. All the qubits, except the $O$ one, have to be measured, specifically in the $\bra{+}$ basis. Above the edges, the terms of the residuals with the interactive coefficients. While the minus sign is justified by Sign Theorem, the $\hat Z$ operator by the Operator Corollary. (c) The same graph, with reduced dimension thanks to the merging of the two nodes in the residual. Thanks to such reduction, the second qubit is measured in the XY plane instead of by $\ket{+}$. As $\ket{o}=\ket{+}$, $\alpha_o=\beta_0=1/\sqrt 2$, and therefore it is possible to apply the opportune byproducts.}
\label{fig:GraphStates}
\end{figure}

For a generic $n$-qubit case, the fully symmetric gauge can be set to be $S^f_{\textbf{i}\textbf{j}} = \bigotimes_{l=1}^{n} S^f_{ij}$.
Consider now the transfer tensor for $n$ qubits in the matrix format. After the fully symmetric gauge has been applied, all the columns are displayed equal to each other:
\begin{equation}
\label{eq:symmetrizedTn}
    T = \frac{1}{2^n}
    \begin{pmatrix}
    T_1 & T_1 & \cdots & T_1 \\
    T_2 & T_2 & \cdots & T_2 \\
    \vdots & \vdots & \ddots & \vdots \\
    T_N & T_N & \cdots & T_N
    \end{pmatrix}
\end{equation}
where $N=2^n$.
Afterward, the $1/2^n$ coefficient can be factorized into $2n$ factors equal to ${1}/{\sqrt 2}$. The output qubits belonging to the generators, whose number amounts to $n$, can be set as $a_i=b_i=1/\sqrt 2$, meaning that the output qubits lie in an overall superposition state $\ ket{+}$. Nevertheless, other $n$ terms survive, which can be addressed as supplementary coefficients $A_i$, $B_i$ to the residual. 
While each interactive term $b_i$ from the generators contributes with the corresponding $\hat X_i$ operator, the $A_i$ ones from the residual, $i\neq 0$, carry a $\hat Z_i$ Pauli matrix. Now the technique which we called \textit{analysis of the residual} can be performed: fitting such coefficients into the transfer tensor in Equation \eqref{eq:symmetrizedTn}, the $T_{i0}$ matrix elements will shape into
\begin{eqnarray}
\label{eq:coeffGauge1}
    T_{00} = T_1 A_1 A_2 ... A_n a_1 a_2 ... a_n \to \hat I \nonumber\\
    T_{10} = T_2 B_1 A_2 ... A_n a_1 a_2 ... a_n \to \hat Z_1 \nonumber\\
    T_{20} = T_3 A_1 B_2 ... A_n a_1 a_2 ... a_n \to \hat Z_2 \nonumber\\
    \vdots& \nonumber\\
    T_{(n-1)0} = T_n B_1 B_2 ... B_n a_1 a_2 ... a_n \to \hat Z_1 ... \hat Z_n
\end{eqnarray}
while the $T_{0j}$ elements will read
\begin{eqnarray}
\label{eq:coeffGauge2}
    T_{00} = T_1 A_1 A_2 ... A_n a_1 a_2 ... a_n \to \hat I \nonumber\\
    T_{01} = T_1 A_1 A_2 ... A_n b_1 a_2 ... a_n \to \hat X_1 \nonumber\\
    T_{02} = T_1 A_1 A_2 ... A_n a_1 b_2 ... a_n \to \hat X_2 \nonumber\\
    \vdots \nonumber\\
    T_{0(n-1)} = T_1 A_1 A_2 ... A_n b_1 b_2 ... b_n \to \hat X_1 ... \hat X_n \nonumber\\
\end{eqnarray}
Thanks to Operator corollary, the $B_i$ interactive terms are linked to the $\hat Z_i$ operators. On the contrary, all the coefficients contained inside the $T_i$ terms are related to qubits not neighboring with the output ones. Referring to Figure \ref{subfig:Rx}, the $a_i$, $b_i$ output coefficients belong to the output qubit $\ket{o}$, the $A_i$, $B_i$ coefficients to the $\ket{B}$ qubit, all the other coefficients in the $T_i$ terms form the generators for the $\ket{a}$ and $\ket{\theta}$ states.

The \textit{analysis of the residual} returns that the combination of the $A_i$, $B_i$ along with the $a_j$, $b_j$ coefficients gives raise to all the $\hat Z_i$ and $\hat X_i$ terms. In order to study the interactions between the qubits in the residual, without any regard to the output, focus on the $T_{00}$ term, as the $\hat Z_i$ operators do not appear. Beside, in order to study the interaction between the residual and the generators, focus on the $T_{1j}$ elements, $j\geq 1$.
From the $T_{00}$ element, in fact, no interactive terms $B_i$, $b_i$ appear, along with no $\hat X_i$, $\hat Z_i$ operators. Thus, it is possible to apply the Sign Theorem, focusing over all the coupled interacting terms $B_k$, $B_l$ contained inside the $T_1$ coefficient, filtering all the other terms with a number $m\neq 2$ of $B_p$ interacting terms.
Resorting back to Sign Theorem, it is possible to check which interactive coefficients from $T_i$ outputs a minus sign in front of the $B_i$. Therefore, all the terms with a number $m\neq 1$ of interactive terms will be filtered away. In such conditions, if two interactive coefficients $B_k$, $B_l$ yield a minus sign in front of the term, the stabilizers of the underlying graph state will reshape as
\begin{equation}
    \langle \hat X_k ; \hat X_l \rangle \to \langle \hat X_k \hat Z_l ; \hat X_l \hat Z_k \rangle
\end{equation}
expressing thus the entanglement between the two qubits in the stabilizer formalism.
A picture representing the coefficients from the $1/2^n$ factor is illustrated in Figure \ref{subfig:FullSymm}. A practical example for a full symmetric gauge transformation is provided for the $\hat R_x$ rotations in Supplementary Note 12.

\section*{\label{sec:12}Supplementary Note 12}

The action of the $\hat R_x(\theta)$ operator, in terms of unitary supplement, can be depicted as
\begin{equation}
\label{eq:Rx}
    \left[ \cos{\left(\frac{\theta}{2}\right)} + i \sin{\left(\frac{\theta}{2}\right)} \hat{X} \right] \left[ a + b \hat{X} \right] \ket{0}
\end{equation}
In the canonical basis, such operation assumes the following matrix form:
\begin{equation}
     \begin{pmatrix}
    a \cos{\left(\frac{\theta}{2}\right)} + i b \sin{\left(\frac{\theta}{2}\right)} &
    b \cos{\left(\frac{\theta}{2}\right)} + i a \sin{\left(\frac{\theta}{2}\right)} \\
    b \cos{\left(\frac{\theta}{2}\right)} + i a \sin{\left(\frac{\theta}{2}\right)} &
    a \cos{\left(\frac{\theta}{2}\right)} + i b \sin{\left(\frac{\theta}{2}\right)}
    \end{pmatrix} \begin{pmatrix}
        1 \\ 0
    \end{pmatrix}
\end{equation}
For sake of simplicity, we now rename $C=\cos{\left(\frac{\theta}{2}\right)}$ and $S= \sin{\left(\frac{\theta}{2}\right)}$. The transfer matrix reads as
\begin{equation}
    T = 
    \begin{pmatrix}
    C a + i S b & C b + i S a \\
    0 & 0
    \end{pmatrix}
\end{equation}
which does not fit the symmetries ruled by the cardinal Equations. The action by the fully symmetric gauge yields
\begin{equation}
    T' = \frac{1}{2}  
    \begin{bmatrix}
    C a + i S b + C b + i S a &
    C a + i S b + C b + i S a \\
    C a + i S b - C b - i S a &
    C a + i S b - C b - i S a 
    \end{bmatrix}
\end{equation}
Applying the rules from Equations \eqref{eq:coeffGauge1} and \eqref{eq:coeffGauge2}, the transfer tensor morphs into
\begin{equation}
\label{eq:RxCoeffs}
\begin{cases}
    T'_{00} =
    (C a + i S b + C b + i S a) A O_p \\
    T'_{01} =
    (C a + i S b + C b + i S a) A O_i \\
    T'_{10} =
    (C a + i S b - C b - i S a) B O_p \\
    T'_{11} =
    (C a + i S b - C b - i S a) B O_i
\end{cases}
\end{equation}
The $A$ and $B$ elements are, respectively, the idle and the interactive coefficients cast into the residual, while $O_p$ and $O_i$ the idle and interactive coefficients for the output qubit. All these new coefficients, $A$, $B$, $O_p$ and $O_i$ are set to $1/\sqrt 2$. We are interested in the sole $T_{00}$ and $T_{21}$ terms, the first one representing the coefficient associated to the $\hat I$ basis element, the second one with the $\hat Z$. As no minus sign occurs between the interacting terms $b$ and $iS$, the respective qubits are not neighbors. On the contrary, a minus sign appears in front of the $Cb$ and $iSa$ terms, suggesting that such qubits neighbor with the $\hat G(A,B)$ one. For construction, the $\hat G(A,B)$ and the $\hat G(O_p,O_i)$ interact among them, and the overall graph state is therefore provided in figure \ref{subfig:Rx}.
Within such scheme, the measurements over all the qubits are given by
\begin{equation}
    \bra{0} [\hat I + \hat X]
\end{equation}
i.e. by a projection over the $\ket{+}$ state.

Nevertheless, it is possible to manipulate Equation \eqref{eq:RxCoeffs} summing over $C+iS =e^{i\theta/2}$:
\begin{equation}
\begin{cases}
    T'_{00} =
    (a e^{i\theta/2} + b e^{i \theta/2}) A O_p \\
    T'_{01} =
    (a e^{i\theta/2} + b e^{i\theta/2}) A O_i \\
    T'_{10} =
    (a e^{-i\theta/2} - e^{-i\theta/2} b) B O_p \\
    T'_{11} =
    (a e^{-i\theta/2} - e^{-i\theta/2} b) B O_i
\end{cases}
\end{equation}
It is possible to trace out an irrelevant global phase $e^{i\theta}$, making the transfer tensor equal to
\begin{equation}
\begin{cases}
    T'_{00} =
    (a + b) A O_p \\
    T'_{01} =
    (a + b) A O_i \\
    T'_{10} =
    (a - b) e^{-i\theta}  B O_p \\
    T'_{11} =
    (a - b) e^{-i\theta}  B O_i
\end{cases}
\end{equation}
Applying the Coefficient Theorem, it is possible to split $e^{-i\theta}$ and $B$ as two interactive coefficients, from a measurement and from a generator respectively, $b$ interacting with $B$ and therefore yielding the minus sign. The overall graph state is now described in Figure \ref{subfig:Rx1}, and the corresponding algebraic description, in terms of generators, is provided by
\begin{equation}
    [a + b \hat X_1 \hat Z_2] [A + B \hat Z_1 \hat X_2 \hat Z_3] [O_p + O_i \hat Z_2 \hat X_3]
\end{equation}
Recalling that $A=B=O_i=O_p=1/\sqrt 2$, the overall result is
\begin{equation}
\label{eq:stabRx}
    \frac{1}{2}[a + b \hat X_1 \hat Z_2] [\hat I + \hat Z_1 \hat X_2 \hat Z_3] [\hat I + \hat Z_2 \hat X_3]
\end{equation}
The first qubit, pinned by $\hat X_1$, is measured in the $\ket{+}$ basis, while the second one, provided by the $\hat Z_1 \hat X_2 \hat Z_3$ generator, is projected into the following state:
\begin{equation}
\label{eq:XYmeas}
    \bra{0} [\hat I + e^{-i\theta} \hat X]
\end{equation}
Equation \eqref{eq:stabRx} is the stabilizer state to be measured, in order to achieve the transformation in Equation \eqref{eq:Rx} (rotation around the $x$-axis) via the fully symmetric gauge.

\section*{Supplementary Note 13}

Beyond the transformations described by the cardinal Equations, another set of gauge operations is provided by the elements from Clifford algebra. As stated in Supplementary Note 3, the elements belonging to the stabilizer group are closed with respect to the Clifford operations. Therefore, Clifford operators can be employed even after the measurements, without circumstancing their action to the sole entanglement of the graph state.
Rotations around the $x$-axis ($\hat R_x$) will suits as an example to deploy. In the first place, we embed the action of a $\hat R_x(\theta)$ gate in the generator formalism:
\begin{equation}
    \hat{R}_x(\theta) \ket{\psi} = [\cos(\theta/2) + \sin(\theta/2) \hat X] [a + b \hat X] \ket{0}
\end{equation}
Of course, such generators do not fit the form of the residual in Equation \eqref{eq:residual2D}. Nevertheless, it is possible to raise a $\hat I = \hat H^2$ operator in front of the kets, so that, employing the rules from Supplementary Note 3, the overall state now reads
\begin{eqnarray}
    [\cos(\theta/2) + \sin(\theta/2) \hat X] [a + b \hat X] \hat H^2 \ket{0} = \nonumber \\
    \hat H [\cos(\theta/2) + \sin(\theta/2) \hat Z] [a + b \hat Z] \frac{1}{\sqrt 2} [\hat I + \hat X] \ket{0}
\end{eqnarray}
making usage of $\hat H \ket{0} = 1/\sqrt 2 [\hat Z + \hat X] \ket{0} = 1/\sqrt 2 [\hat I + \hat X] \ket{0}$. The actual form now fits a possible residual. Reversing the Coefficient Theorem, the second term and the third one are reshaped into
\begin{equation}
\begin{split}
    [\cos(\theta/2) + \sin(\theta/2) \hat Z] \to \bra{YZ(\theta)}_1 \frac{1}{\sqrt 2} [\hat I + \hat X_1 \hat Z] \ket{0} \\
    [a + b \hat Z] \to \bra{XY(0)}_2 [a + b \hat X_2 \hat Z] \ket{0} \\
    \frac{1}{\sqrt 2} [\hat I + \hat X] \ket{0} \to \frac{1}{\sqrt 2} [\hat I + \hat X \hat Z_1 \hat Z_2] \ket{0} \nonumber\\
\end{split}
\end{equation}
$\bra{YZ(\theta)}_1$ and $\bra{XY(0)}_2$ belong to the measurement projections described in Supplementary Note 4, and the indices underlines which space the $\hat X_i$ operators in Equation \eqref{eq:measurements} act on.
The graph state to be measured displays now as
\begin{equation}
    \frac{1}{2} [\hat I + \hat X_1 \hat Z] [\hat I + \hat X \hat Z_1 \hat Z_2] [a + b \hat X_2 \hat Z] \ket{000}
\end{equation}
After the $\bra{YZ(\theta)}_1 \,$, $\bra{XY(0)}_2$ measurements, the $\hat H$ Hadamard operator needs to be applied on the output state.

\section*{Supplementary Note 14}

The gauge freedom summarized by the cardinal equations makes straightforward to implement an equivalence relation between different graphs:
\begin{equation}
\begin{split}
    G[V_1,E_1] \sim_{U} G[V_2, E_2] \Leftrightarrow \exists \, S : \mathfrak{T} \to \mathfrak{T} \, | \, T^1_{\textbf{i}} = S_{\textbf{i}\textbf{j}} T^2_{\textbf{j}} \\
    \lor \, T^2_{\textbf{i}} = S_{\textbf{i}\textbf{j}} T^1_{\textbf{j}}, \\
    T^1_{\textbf{i}} T^1_{\textbf{j}} = T^1_{\textbf{i}\circ \textbf{j}} T^1_{\textbf{j} \circ \textbf{j}}, \;
    T^2_{\textbf{i}} T^2_{\textbf{j}} = T^2_{\textbf{i}\circ \textbf{j}} T^2_{\textbf{j} \circ \textbf{j}}
\end{split}
\end{equation}
Here $G$ represent a simple graph, i.e. a graph with no loops neither multiple edges~\cite{west2001introduction}. The sole distinction between these two $G$ graphs is remarked by the sets of vertices and edges $[V_i, E_i]$. The equivalence relation $\sim_U$ is given with respect to a certain unitary operator $\hat U$, meaning that the MBQC operations over $G[V_1,E_1]$ and $G[V_2, E_2]$ will yield the same outcome. $\mathfrak{T}$ stands for the space $[\mathbb{C}^{4}]^{\otimes r}$, $r$ being the rank of $T^1$ and $T^2$, the transfer tensors underlying the first and the second graphs. However, due to the potential singularity of the $S$ tensor, it is mandatory to state that the $S$ gauge operator could both map $T_1 \to T_2$ or $T_2 \to T1$. The last constraint is pointed out by the $T_{\textbf{i}} T_{\textbf{j}} = T_{\textbf{i}\circ \textbf{j}} T_{\textbf{j} \circ \textbf{j}}$, as such condition endows $T$ with the required symmetry to represent a graph in the stabilizer formalism.
Once a unitary operator $\hat U$ can be implemented by a set of measurements over a graph $G[V,E]$, it follows that $G \in [U]$, i.e. $G$ belongs to the class of graphs able to operate the $\hat U$ unitary transformation, thanks to the measurement techniques described in the MBQC theory.

\section*{Supplementary Note 15}

\begin{figure}
\subfloat[\label{subfig:QFT}]{{\includegraphics[width=0.9\textwidth]{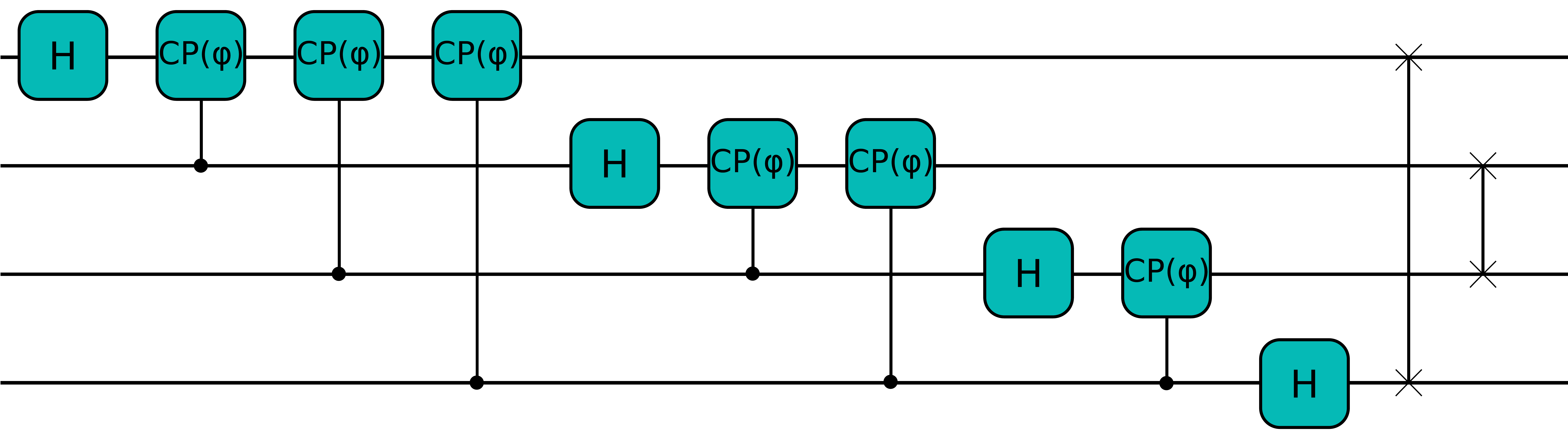} }}
\\
\subfloat[\label{subfig:CPhase}]{{\includegraphics[width=0.9\textwidth]{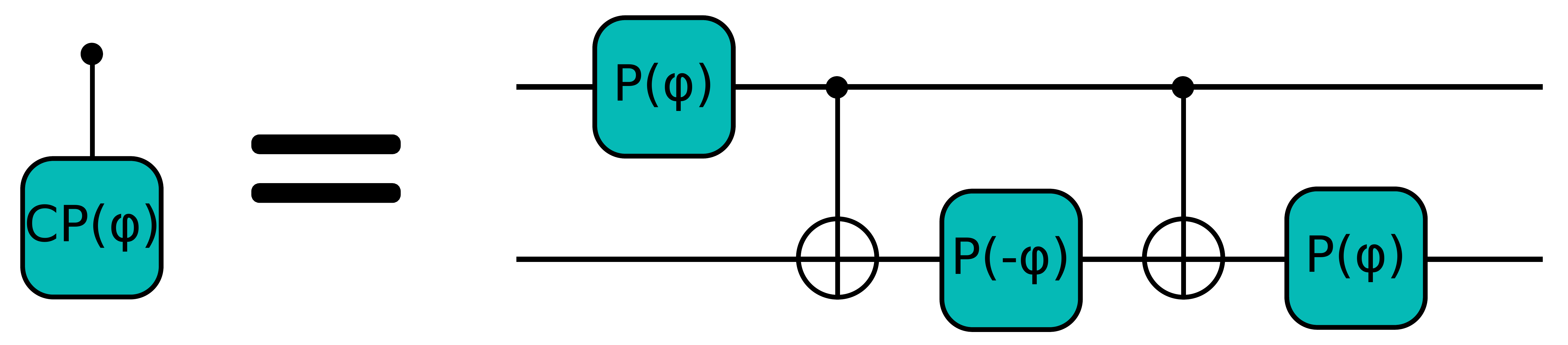} }}
\\
\subfloat[\label{subfig:Cphase_Mcalc}]{{\includegraphics[width=0.8\textwidth]{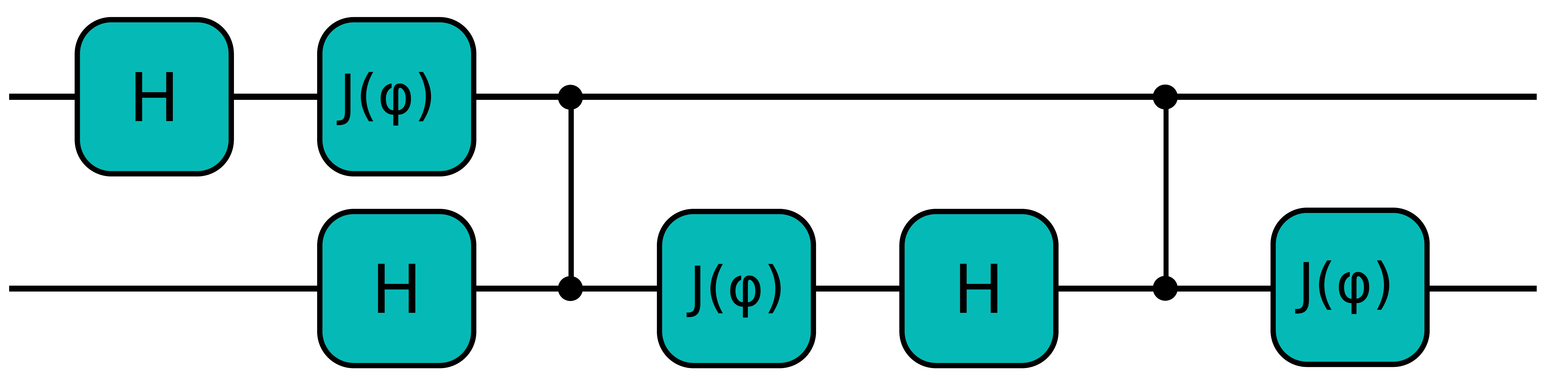} }}
\caption{(a) The scheme for the Quantum Fourier Transform implemented with $4$ qubits. (b) The controlled phase is decomposed in terms of Clifford gates and rotations. (c) Decomposition of the controlled-phase gate in terms of $\hat J(\phi)$ and $\hat H$ operators, $\hat H$ being $\hat J(0)$}
\label{fig:QFT}
\end{figure}

Here the number of source qubits involved by both the M-Calculus and the fully symmetric gauge are compared.
By invoking the rules from the M-Calculus, the total number of qubits is left invariant for both optimized and unoptimized CME patterns. From the Supplementary Note \ref{sec:12}, instead, the number of qubits $N_{fully}$ is explained to scale as
\begin{equation}
    N_{fully} = n_{inputs} + n_{operations} + 2n_{inputs} = 3n_{inputs} + n_{operations}
\end{equation}
\textbf{Example 1: Quantum Fourier transform}
In order to provide an explicit example, a paramount and widespread algorithm is considered, consisting of the quantum Fourier transform. The quantum Fourier transform involves the applications of several controlled phase gates, whose number increases along with the number $n$ of involved qubits. First, a decomposition of such controlled rotation gate is required. A valid model is provided in Figure \ref{subfig:CPhase}. Such circuit must therefore be converted into $\hat J(\phi)$ operators in order to perform the M-Calculus, which is exposed in Supplementary Figure \ref{subfig:Cphase_Mcalc}. The conversion is straightforward, once we acknowledge the fact that
\begin{equation}
    \hat P(\phi) = \hat H \hat J(\phi)
\end{equation}
On the other hand, the fully symmetric gauge requires a decomposition in terms of Pauli matrices. By applying the rules from the Clifford group, the controlled-phase gate turns out to be
\begin{equation}
    \hat{CP}(\phi) = [\cos(\phi) + i \sin(\phi) \hat Z_1] [\cos(\phi) + i \sin(-\phi) \hat Z_1 \hat Z_2] [\cos(\phi) + i \sin(\phi) \hat Z_2]
\end{equation}
Approaching the M-Calculus, therefore, as $\hat H = \hat J(0)$, a qubit to be measured is involved per each Hadamard gate, whose number we refer to as $n_H$. Per each controlled-phase gate instead, as from Figure \ref{subfig:Cphase_Mcalc}, six measurements arise. Moreover, an additional qubit is required for any input qubit, whose total amount we refer to as $n$.
Afterwards, the final swaps need to be taken into account. The number of swap operations $n_{swap}$ scales as $\lfloor\frac{n}{2}\rfloor$ with respect with the total amount of qubits $n$, thus, in an asymptotic regime, as $n/2$. The swap gate can be decomposed into three CNOT gates:
\begin{equation}
    SWAP_{ij} = \hat{CX}_{ij} \hat{CX}_{ji} \hat{CX}_{ij}
\end{equation}
Due to the decomposition $\hat{CX}_{ij} = \hat H_j \hat{CZ} \hat H_j$, the number of measurements in the M-Calculus amounts to $6$ per swap. On the contrary, in the fully symmetric gauge, the CNOT gates can backpropagate at the head of the circuit, where $\hat{CX}_{ij} \ket{0}^{\otimes n}$, transforming all of the $\hat K_i$, $\hat K_j$ stabilizers but without affecting the number of required qubits.
The overall scaling for the number of measurements $N^{Mc}_{op}$, varying with the number of input qubits $n$, is reported to be
\begin{equation}
    N_{op}^{Mc} = n + n_H + n_{CP} \frac{n(n-1)}{2} + 6n_{swap} = 5n + 3 n(n-1)
\end{equation}
where $6n_{swap} = 6n/2=3n$, while $n_H=1$.
The $n(n-1)$ factor is linked to the number of controlled-phase operations, scaling from $1$ to $n$ for each qubit to be added. The overall summation, $\sum_{i=1}^n i$, returns indeed $n(n-1)/2$. Furthermore, the number of Hadamard gates $n_H$, as previously states, equals the total number $n$ of qubits.
Switching to the fully symmetric gauge, the total number of operations amounts to $n$ for the Hamadard gates, $3 n_{CP}$ for the controlled-phase rotations and $2n$ due to the peculiar structure of the gauge:
\begin{equation}
    N_{op}^{fully} = 3n + \frac{3}{2} n(n-1)
\end{equation}
Summing up the two scaling laws, we can eventually benchmark their complexity in terms of measurements and therefore involved qubits. As both the laws scale asymptotically with $n^2$, it is straightforward to check that the M-calculus requires double the size of the graph state, $O(3 n^2)$, compared to the fully symmetric gauge, $O(3 n^2/2)$.

\textbf{Example 2: QAOA}
To carry a new example, we now focus on the QAOA, a variational algorithm whose depth can scale at will with the number $p$ of layers. When the QAOA is adapted to tackle the Max-Cut problem, each layer is composed by a set of two operations, i.e. a two-qubits rotation $\hat R_{z z}$ and single-qubit rotations $\hat R_x$. In the M-calculus frame, the $\hat R_x$ rotations demand three qubits to be implemented:
\begin{equation}
    \hat R_x\left(\frac{\theta}{2}\right) = \hat J(\theta) \hat H = X_3^s M_2^{-\theta} E_{2,3} X_2^s M_1^{0} E_{1,2}
\end{equation}
On the other hand, each $R_{zz}$ operation can be decomposed as follows:
\begin{equation}
    \hat R_{z_i z_j}(\theta) = [\cos(\theta) + i \sin(\theta) \hat Z_i \hat Z_j] =
    \hat{CX}_{ij} [\cos(\theta) + i \sin(\theta) \hat Z_i] \hat{CX}_{ij} = 
    \hat H \hat{CZ}_{ij} \hat H \hat R_{z_i}(\theta) \hat H \hat{CZ}_{ij} \hat H_{ij}
\end{equation}
from which $4$ qubits are required by the CME decomposition, plus two additional input qubits.
The fully symmetric gauge, instead, enables to convert directly the $[\cos(\theta) + i \sin(\theta) \hat Z_i \hat Z_j]$ and $[\cos(\theta) + i \sin(\theta) \hat X_i]$ gates, already decomposed in terms of Pauli matrices.
The number of $\hat R_{zz}$ operations, when approaching the Max-Cut problem, depends on the total amount $\mathcal{E}$ of connections in the graph it has to be partitioned. We consider two limit cases, i.e. the cyclic graph and the complete graph. In the first case, $\mathcal{E} = n$, in the latter $\mathcal{E} = n(n-1)/2$. Hereafter, we can benchmark the complexity between the M-calculus and the fully symmetric gauge for the cyclic case:
\begin{equation}
\begin{cases}
    N_{op}^{Mc} = n + p(4n + 3n) = n(1+7p) \\
    N_{op}^{fully} = 2n + 2np = 2n (1+ p)
\end{cases}
\end{equation}
Even for a number of layers $p=1$, the M-calculus requires the double of qubits, while for $p\gg 1$ the ratio asymptotically scales to $7/2$. We now revert to the complete graph:
\begin{equation}
\begin{cases}
    N_{op}^{Mc} = n + 2pn(n-1) + 3pn \\
    N_{op}^{fully} = 2n + \frac{1}{2}pn(n-1) + np
\end{cases}
\end{equation}
In such case, the leading term is $n^2$, and for $n\gg1$ the M-calculus requires $4$ times the number of qubits with respect to the fully symmetric gauge.

\bibliography{notes}

\end{document}